\documentclass[journal]{IEEEtran}

\ifCLASSINFOpdf
\else
\fi
\usepackage{caption}
\usepackage{amsmath}
\usepackage{amsthm}
\usepackage{amssymb}

\usepackage{graphicx}
\usepackage{mathtools}
\usepackage{multicol}
\usepackage[justification=centering]{caption}
\usepackage{autobreak}
\allowdisplaybreaks[0]
\usepackage{caption}
\usepackage{graphicx}
\usepackage{float} 
\usepackage{subcaption}
\usepackage{stfloats}
\usepackage{cite}
\usepackage{xcolor}
\usepackage{graphicx}
\allowdisplaybreaks 

\begin{document}

%
\title{Network-Wide PAoI Guarantee in CF-mMIMO Networks with S\&C Coexistence: A Unified Framework for Spatial Partitioning \\Toward xURLLC}

%
%
%
 \author{Yanxi~Zhang,~\IEEEmembership{Student Member,~IEEE,} Mingwu~Yao,~\IEEEmembership{Member,~IEEE,} Qinghai~Yang,~\IEEEmembership{Member,~IEEE} and Muyu~Mei 
 \vspace{-1em}
 
 \thanks{Yanxi Zhang, Mingwu Yao, and Qinghai Yang are with the State Key Laboratory of Integrated Services Networks, School of the Telecommunication Engineering, Xidian University, Xi’an 710071, China (e-mail: yanxi.zhang@stu.xidian.edu.cn; mwyao@xidian.edu.cn; qhyang@xidian.edu.\\cn). Muyu Mei is with the School of Computer Science and Communication Engineering, Jiangsu University, Zhenjiang 212013, China (e-mail: mei\_muyu@163.com).}
 \vspace{-1em}
}
\maketitle

\begin{abstract}
As a key capability of 6G, sensing-communication (S\&C) coexistence over distributed infrastructure is expected to support next-generation ultra-reliable and low-latency communication (xURLLC) applications, which demand both robust connectivity and real-time environmental awareness. This paper investigates network-wide information freshness in large-scale cell-free massive multiple-input multiple-output (CF-mMIMO) with S\&C coexistence. A challenge arises from the spatial partitioning of access points (APs) into S\&C roles: allocating more APs to sensing improves update generation, whereas allocating more APs to communication enhances reliable short-packet delivery. To address this, we develop a unified analytical framework by combining stochastic geometry and stochastic network calculus (SNC) to characterize the peak age of information (PAoI) violation probability (PAVP). Specifically, we derive the moment generating functions (MGFs) of sensory packet inter-arrival and service times, accounting for the joint stochastic spatial distribution of APs and users, imperfect channel state information (CSI), and finite blocklength coding (FBC). This facilitates the derivation of a tractable upper bound on the PAVP, which is minimized to determine the optimal AP partitioning. The derived bound accurately captures the performance trend and yields a minimizing partition factor that closely matches simulations. Therefore, the framework provides an efficient and low-complexity tool for network-wide PAoI guarantee and coexistence-oriented design in CF-mMIMO networks toward xURLLC.
\end{abstract}


%
\IEEEpeerreviewmaketitle

\vspace{-1em}
\section{Introduction}
\vspace{-0.5em}
%
%
%
%
\IEEEPARstart{T}{he} forthcoming sixth-generation (6G) wireless networks are envisioned to evolve beyond conventional data delivery into intelligent infrastructures that seamlessly support communication and environmental awareness \cite{ref1}. This dual requirement is particularly important for next-generation ultra-reliable and low-latency communication (xURLLC) applications, such as factory automation, autonomous driving, and remote surgery \cite{ref3}, where both delayed delivery and stale situational information may cause severe performance degradation. In large-scale distributed deployments, a practical way to support this dual functionality is through sensing-communication (S\&C) coexistence over a common wireless infrastructure. In such architectures, sensing and communication may operate orthogonally while still being tightly coupled through shared access-point deployment and resource budgets.

Conventional performance metrics, such as latency and reliability, are often insufficient for capturing the strict timeliness requirements of these services \cite{ref4}. As a result, the age of information (AoI) has emerged as an essential metric for quantifying data freshness from the user's perspective, measuring the time elapsed since the generation of the most recently received update \cite{ref5}. For xURLLC applications, decisions based on stale information can lead to catastrophic failures. Thus, controlling the peak AoI (PAoI), which reflects worst-case data staleness, is of particular importance. A network-wide PAoI guarantee is thus a meaningful objective for coexistence-oriented wireless systems serving time-sensitive applications.

To meet such stringent demands across wide areas, the cell-free massive multiple-input multiple-output (CF-mMIMO) architecture presents a compelling network foundation. By enabling a large number of distributed access points (APs) to cooperatively serve users without cell boundaries, CF-mMIMO mitigates inter-cell interference, improves macro-diversity, and delivers more uniform service quality \cite{ref6, ref7}. These properties make CF-mMIMO particularly attractive for coexistence scenarios in which both sensing coverage and communication reliability depend on the same distributed infrastructure. In this sense, the benefit of CF-mMIMO in the present work lies not in waveform-level integration, but in providing a scalable spatial substrate on which sensing and communication coexist and compete for network resources.

Despite these advantages, analyzing AoI in large-scale CF-mMIMO coexistence networks remains highly challenging. In practical deployments, AP locations are often irregular due to site availability and terrain constraints, while user locations are inherently random and uncoordinated. This double spatial randomness interacts with small-scale fading, channel estimation uncertainty, finite blocklength transmission, and queuing dynamics, leading to intricate statistical dependencies across the network. These effects become especially pronounced in xURLLC scenarios, where stringent reliability and latency requirements exacerbate the challenge of maintaining timely information over the entire spatial domain. Thus, any realistic performance guarantee must be statistical and network-wide, averaging over random topologies and channel realizations. 

Beyond these analytical challenges, a central design issue in the considered architecture is how to allocate shared network resources between sensing and communication \cite{ref8}. Under a fixed AP density and bandwidth budget, assigning more APs to sensing improves sensing coverage and accelerates update generation, but leaves fewer communication resources for reliable low-latency packet delivery.  Conversely, allocating more APs to communication strengthens the transmission infrastructure while reducing the sensing capability that drives timely information generation. This tension constitutes a critical bottleneck for maximizing network-wide information freshness \cite{ref9}. Despite its significance, the impact of spatial resource partitioning on network-level AoI performance lacks a comprehensive theoretical characterization. Therefore, a unified analytical framework is required to establish the fundamental principles governing this interplay in large-scale distributed CF-mMIMO systems.

This paper addresses the above gap by developing a tractable analytical framework for evaluating and optimizing network-wide information freshness in large-scale CF-mMIMO networks with S\&C coexistence. The core of our contribution is the derivation of a tractable network-wide upper bound on the PAoI violation probability (PAVP). This bound analytically links the statistical deployment of distributed network resources to key application-layer metrics. The utility of the framework is demonstrated through an analysis of the derived bound, which unveils the network-level S\&C trade-off and provides theory-grounded insights for system design. The main contributions of this work are summarized as follows:
\begin{itemize}
\item We establish a network-level coexistence framework for large-scale CF-mMIMO networks supporting xURLLC services. The spatial distributions of APs and users are modeled as independent poisson point processes (PPPs). And a spatial partitioning scheme is introduced to allocate APs to either sensing or communication, thereby enabling network-wide freshness analysis under shared infrastructure constraints.

\item Based on this model, we derive tractable expressions for (i) the network sensing coverage probability and the corresponding moment generating function (MGF) of sensory packet inter-arrival times, and (ii) the network communication coverage probability and MGF of packet service times, both explicitly accounting for topological randomness and channel dynamics.

\item We develop a unified analytical framework using SNC to relate sensing and communication performance to end‑to‑end information freshness. By systematically integrating the MGFs of packet inter-arrival and service times, we derive a tractable upper bound on the network-wide PAVP. This bound formulates an optimization problem that determines the spatial AP partitioning required to minimize PAVP, thereby analytically resolving the network-level S\&C trade-off.

\item We conduct extensive simulations to validate the proposed framework. The results show that the analytical bound accurately preserves the performance trend and the minimizing partition factor. This makes the derived bound an efficient low-complexity surrogate for network-wide PAoI guarantee and coexistence-oriented system design.

\end{itemize}

The proposed framework is intended primarily for network planning and slow-timescale coexistence adaptation, where low-complexity optimization is essential and exhaustive simulation-based tuning is impractical.

\vspace{-1.2em}

\subsection{Related Works}
\vspace{-0.6em}
Our work is theoretically grounded in the definition and analysis of AoI. Since its inception \cite{ref10}, AoI has been extensively studied in various queuing systems with a primary focus on average AoI (AAoI) and PAoI \cite{ref11, ref12}. These studies established fundamental principles for freshness analysis under different scheduling disciplines and traffic models \cite{ref13, ref14}. More recently, the focus has shifted from average metrics to tail-oriented guarantees, which are particularly relevant to xURLLC services. In this context, SNC has emerged as a powerful tool for deriving statistical tail bounds on the AoI violation probabilities in stochastic queuing systems \cite{ref15, ref16}. However, most existing AoI studies assume exogenous update arrivals, where packet generation is independent of the sensing process and the underlying wireless infrastructure.

Simultaneously, the S\&C coexistence and coordination in wireless networks have attracted significant interest \cite{ref17, ref18, ref19}. Existing studies have shown that S\&C functions often compete for shared resources, giving rise to design trade-offs\cite{ref20, ref21, ref22, ref23}. Although recent works have considered freshness-aware S\&C systems \cite{ref24, ref25}, they mainly focus on small-scale settings or abstract queuing formulations and do not capture the network-wide interaction between sensing-side update generation and communication-side packet delivery over a random large-scale infrastructure. In such networks, the spatial partitioning of resources between S\&C constitutes a dominant yet unexplored design dimension for ensuring freshness.

Meeting stringent AoI requirements in xURLLC regimes requires physical layer modeling that incorporates FBC \cite{ref5}, as this accurately captures the rate-reliability trade-off in short-packet communications. Achieving the extreme reliability levels necessitated by xURLLC often further requires retransmission mechanisms \cite{ref26, ref27}. While prior works have studied the impact of FBC and retransmissions on delay and link reliability \cite{ref28, ref29}, they typically assume fixed topologies, conventional cellular structures, or exogenous traffic processes. Their integration with sensing-driven arrivals, stochastic topology, and CF-mMIMO-based cooperative transmission remains largely unexplored from an information freshness perspective.

To realize S\&C at scale and support xURLLC applications, the CF-mMIMO architecture has emerged as a promising enabler due to its ability to provide uniform coverage through distributed cooperation \cite{ref30, ref31}. Stochastic geometry has become a standard tool for analyzing such large-scale networks, with extensive research on metrics like coverage probability \cite{ref32, ref33} and ergodic rate \cite{ref34, ref35} under practical impairments such as pilot contamination \cite{ref36, ref37}. While communication performance analysis for CF-mMIMO is well established, its integration with S\&C regarding information freshness remains in its infancy. To the best of our knowledge, no prior work has addressed the fundamental spatial partition problem in large-scale cell-free S\&C coexistence networks or quantified the resulting impact on AoI while accounting for FBC and retransmissions.

Despite substantial progress in AoI analysis, coexistence design, xURLLC optimization, and large-scale network modeling, a gap remains at their intersection. The literature still lacks a tractable analytical framework for quantifying network-wide PAoI in large-scale CF-mMIMO coexistence networks, where sensing-driven packet generation and communication-limited packet delivery are jointly shaped by stochastic topology and shared infrastructure constraints. This paper aims to bridge this gap by developing an SNC-based analytical framework for evaluating and optimizing PAVP under spatial AP partitioning. Importantly, the resulting bound is used not only as a conservative performance guarantee, but also as a low-complexity surrogate for identifying the freshness-optimal coexistence configuration.

\vspace{-0.8em}
\subsection{Organizations}
\vspace{-0.1em}
The remainder of this paper is organized as follows. Sec. II describes the system model and formulates the core optimization problem. Sec. III presents the SNC-based analytical framework for PAVP analysis. Sec. IV investigates the sensing subsystem to derive packet inter-arrival time statistics. Sec. V and VI focus on the communication subsystem: Sec. V addresses uplink channel estimation, while Sec. VI analyzes downlink transmission to derive the packet service time statistics. Sec. VII integrates these results to evaluate and optimize the network-wide PAVP. Numerical results are presented in Sec. VIII to validate our framework and offer design insights. Finally, Sec. IX concludes the paper.
\vspace{-0.8em}
\subsection{Notations}
\vspace{-0.1em}
Scalars, vectors, and matrices are denoted by italic letters, boldface lowercase letters, and boldface uppercase letters, respectively. The superscripts $(\cdot)^T$, $(\cdot)^*$, and $(\cdot)^H$ denote the transpose, complex conjugate, and Hermitian transpose operations. The operators $\operatorname{tr}(\cdot)$ and $\operatorname{diag}(\cdot)$ represent the matrix trace and the diagonal matrix formation, respectively.

\vspace{-0.7em}
\section{System Model and Problem Statement}
\vspace{-0.1em}
\subsection{Network Model}
\begin{figure}[t]
    \centering{
    \includegraphics[width=8.85cm]{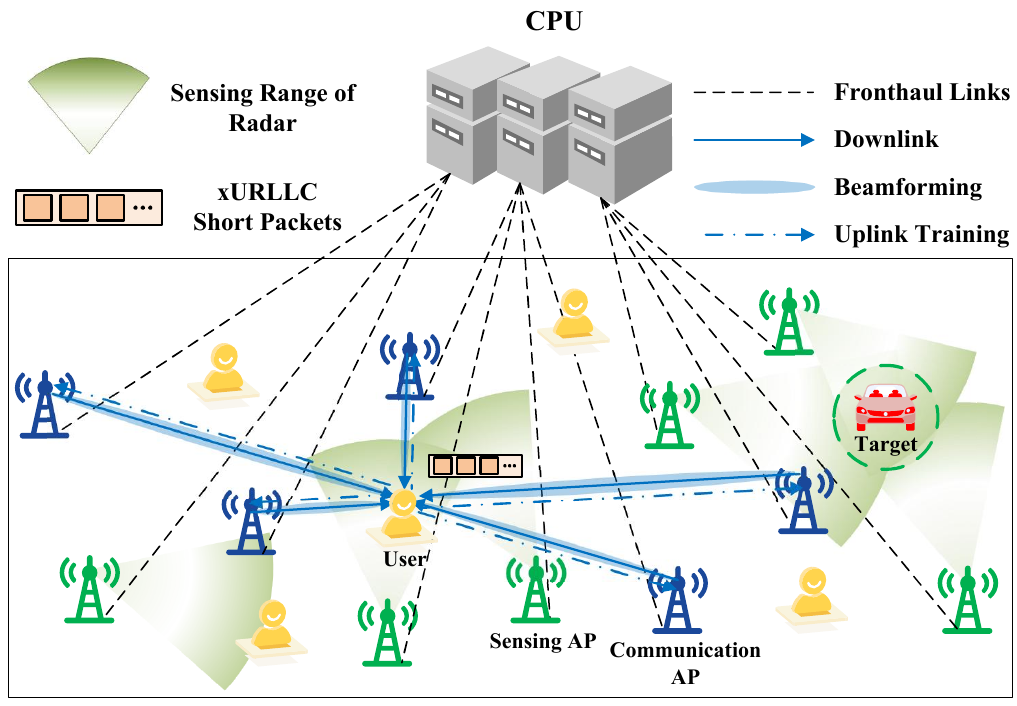}
    \captionsetup{justification=raggedright,singlelinecheck=false} 
    \caption{The CF-mMIMO coexistence networks.}
    \label{fig:1}}
    \vspace{-1.5em}
\end{figure}
As illustrated in Fig. 1, we consider a S\&C coexistence framework for emerging xURLLC applications within a CF-mMIMO architecture. The network comprises dual-functional APs distributed according to a two-dimensional homogeneous PPP $\Phi$ with intensity $\lambda$. Each AP operates exclusively as either a monostatic radar for sensing or a transceiver for communication within the considered large-timescale partitioning horizon. This functional partitioning is governed by a parameter $\beta \in [0, 1]$, which thins the original PPP into two independent PPPs: the sensing AP set $\Phi_s$ with intensity $\lambda_s = (1-\beta)\lambda$ and the communication AP set $\Phi_c$ with intensity $\lambda_c = \beta \lambda$. All APs connect to a central processing unit (CPU) via ideal fronthaul links\footnote{This assumption is adopted to isolate the radio-access contribution to AoI; non-ideal fronthaul latency can be incorporated in the SNC framework as an additional service component, which is left for future work.} and operate with uniform power $P$. 

Sensing and communication functionalities are operated orthogonally via frequency-division multiplexing to eliminate inter-functional interference. The total system bandwidth $B$ is partitioned into $B_s$ and $B_c$ for sensing and communication ($B = B_s + B_c$), while intra-functional interference remains within each set. Furthermore, the communication functionality employs time-division duplexing (TDD), where each coherence interval comprises an uplink pilot training phase followed by a downlink data transmission phase.

In the CF-mMIMO paradigm, communication APs in $\Phi_c$ are equipped with $N$ antennas and collaboratively serve users distributed according to an independent PPP $\Phi_u$ with intensity $\lambda_u$. We assume a massive MIMO regime where the effective transmit antenna density significantly exceeds user density. For analytical tractability, we focus on a typical user $k$ at the origin. By Slivnyak's theorem \cite{ref38}, this user's performance is statistically representative of the spatial average.

The study examines the information flow from target detection to user reception. As the network monitors a randomly located target, APs in $\Phi_s$ perform periodic scans at intervals $T_s$. A network-wide sensing event is successful if at least one AP detects the target, triggering the generation of a short sensory packet of fixed length $L \leq 100$ bits, indexed by $n \in \mathbb{N}$. Containing time-critical updates, these packets are forwarded to the CPU and queued in a first-come-first-served (FCFS) buffer. The head-of-line packet is then cooperatively transmitted by the APs in $\Phi_c$ to the user. To model the short-packet transmissions essential for xURLLC, we use FBC theory \cite{ref5} to approximate the achievable rate as
\begin{align}
R\left(\gamma\right) \approx \frac{B_c}{\ln 2}\left[\ln \left(1+\gamma\right)\!-\!\sqrt{\frac{V\left(\gamma\right)}{\mathcal{N}}} Q^{-1}\left(\epsilon\right)\right], 
\label{a}
\end{align}
\noindent where $\epsilon$ is the decoding error probability (DEP), $\mathcal{N}$ is the blocklength, $\Gamma$ is the random signal-to-interference-plus-noise ratio (SINR), with $\gamma$ representing its specific realization, $Q^{-1}(\cdot)$ is the inverse Gaussian Q-function, and the channel dispersion is $V\left(\gamma\right)=1-{\left(1+\gamma\right)^{-2}}.$ To satisfy the stringent reliability requirements of xURLLC, the receiver provides immediate ACK/NACK feedback, and incorrectly decoded packets are retransmitted until successful reception.
\subsection{Channel Model}
We adopt a block fading channel model for communication, where channel coefficients remain static within a coherence interval $\tau_c$ but vary independently across intervals. Each interval includes an uplink training phase of $\tau_{tr}$ symbols for CSI acquisition and a downlink data transmission phase of $\tau_d$ symbols. TDD operation exploits channel reciprocity, allowing CSI estimated from uplink pilots to be used for downlink precoding. For a given realization of $M$ APs, the channel vector between the $m$-th AP ($m=1, \ldots, M$) and user $k$ is
$$
\mathbf{h}_{m k}=l_{m k}^{1 / 2} \mathbf{g}_{m k},
$$
where $l_{mk}$ is the large-scale fading coefficient and $\mathbf{g}_{mk} \in \mathbb{C}^{N \times 1}$ is the small-scale fading vector. The coefficient $l_{mk}$ follows a bounded path-loss model based on distance $r_{mk}$:
$$
l_{m k}\left(r_{m k}\right)=\min \left(1, r_{m k}^{-\alpha}\right),
$$
with path-loss exponent $\alpha > 2$. This model prevents infinite channel gain singularities as $r_{mk} \to 0$. The small-scale fading vector $\mathbf{g}_{mk}$ follows a Rayleigh fading distribution with independent and identically distributed (i.i.d.) elements, such that $\mathbf{g}_{mk} \sim \mathcal{CN}(\mathbf{0}, \mathbf{I}_N)$.
\subsection{Performance Metric and Problem Statement}
In mission-critical xURLLC applications, information timeliness is paramount. Data packets must not only meet reliability requirements but also maintain freshness to enable instantaneous decision-making. We quantify this using the AoI, specifically focusing on the PAoI to capture worst-case data staleness. For the $n$-th sensory packet, the PAoI $\Delta(n)$ is defined as the time elapsed from its generation until the successful reception of the subsequent packet. 

To ensure the PAoI remains below a threshold $\zeta$, we adopt the PAVP, defined as $\mathbb{P}\left[\Delta(n)>\zeta\right]$, as the key performance metric. Minimizing PAVP is essential for xURLLC to ensure decisions rely on timely information, thereby preventing catastrophic failures caused by data staleness. Our primary objective is to determine the optimal partitioning factor $\beta^*$ that minimizes the statistical network-wide PAVP, stated as
\begin{align}
\mathcal{P}: \quad \min _{0 \leq \beta \leq 1}  \quad \mathbb{P}\left[\Delta(n)>\zeta\right].
\label{eq:problem_formulation_rho}
\end{align}

Solving $\mathcal{P}$ yields fundamental insights into how spatial resource partitioning between sensing and communication impacts network-wide information freshness in large-scale, distributed S\&C coexistence networks.
\section{An SNC-based Framework for PAVP Analysis}
Directly computing the PAVP for problem $\mathcal{P}$ is analytically intractable due to the complex interplay between the spatial randomness of the network topology and the temporal stochasticity of channel fading and queuing dynamics. To address this, we employ the SNC framework to derive a tractable upper bound on the PAVP. This approach bridges physical-layer dynamics and application-layer timeliness, facilitating both theoretical analysis and system-level optimization.

\subsection{Queuing Model in the SNC Framework}
The CF-mMIMO coexistence network is modeled as a packet-level queuing system. Let $T^A(n)$ and $T^D(n)$ denote the arrival and departure times of the $n$-th packet, respectively, with $T^A(0) = T^D(0) = 0$. The inter-arrival time between packets $v$ and $n$ is $T^A(v, n) = T^A(n) - T^A(v)$, and the corresponding departure interval is $T^D(v, n) = T^D(n) - T^D(v)$. The instantaneous AoI at time $t$ is defined as 
$$\Delta(t)=t-\max_{n \geq 1}\left\{T^A(n): T^D(n)<t\right\}.$$ Consequently, the PAoI of packet $n$ can be expressed as
\begin{align}
    \Delta(n)&=T^D(n+1)-T^A(n) \nonumber\\
    &=T^D(n+1)-T^A(n+1)+T^A(n,n+1).
    \label{c}
\end{align}
As the PAoI depends on arrival and departure timestamps, we adopt a packet-based SNC formulation grounded in the following definition and lemmas.

\textit{Definition 1:} A system with arrival process $T^A(n)$ and departure process $T^D(n)$ is an exact $T^S(v, n)$ server if, for all $n \in \mathbb{N}^+$, the departure times satisfy \cite{refZhang}:
$$
T^D(n)=\max _{1 \leq v \leq n}\{T^A(v)+T^S(v, n)\}.
$$

\textit{Lemma 1:} An FCFS system with packet service times $T^S(n)$ acts as an exact server with service process $T^S(v, n)=\sum_{\ell=v}^n T^S(\ell)$. The departure time for any packet $n \geq 1$ is
\begin{align}
T^D(n)=\max _{1 \leq v \leq n}\left\{T^A(v)+\sum_{\ell=v}^n T^S(\ell)\right\}.
\label{l1}
\end{align}

\textit{Proof:} Let $\mathcal{Z}(n)$ and $\mathcal{L}(n)$ denote the service start time and duration for packet $n$, respectively. In an FCFS system, a packet commences service only after it has arrived and the preceding packet has departed. Thus, for $n \geq 2$, $\mathcal{Z}(n)=\max \left\{T^A(n), \mathcal{Z}(n-1)+T^S(n-1)\right\},$ with $\mathcal{Z}(1)=T^A(1)$. Recursively expanding this yields the service start time:
\begin{align}
\mathcal{Z}(n)=\max _{1 \leq v \leq n}\left\{T^A(v)+\sum_{\ell=v}^{n-1} T^S(\ell)\right\}. 
\label{l2}
\end{align}
The departure time is $T^D(n)=\mathcal{Z}(n)+\mathcal{L}(n)$. Substituting Eq. \eqref{l2} into this relationship yields Eq. \eqref{l1}, confirming that the FCFS system behaves as an exact server with an additive service process. \hspace*{\fill}\qed

Substituting \textit{Definition 1} into Eq. \eqref{c}, the PAoI can be reformulated as
\small
\begin{align}
    \Delta(n)\! =\! \max _{\!1 \leq v \leq n\!+\!1\!}\{T^S\!(v, n\!+\!1)\!-\!T^A\!(v, n\!+\!1)\} \!+\!T^A\!(n,n\!+\!1).
    \label{d}
\end{align}
\normalsize
Since direct PAVP computation remains infeasible, we use the MGF bounding technique. The MGF of a non-negative random variable $X$ is $\mathcal{M}_X(\theta)=\mathbb{E}\left[e^{\theta X}\right]$ for $\theta > 0$. The following lemma provides an MGF-based upper bound for the PAVP.

\textit{Lemma 2:} For an FCFS system satisfying the stability condition $\mathcal{M}_{T^A(n, n+1)}(-\theta) \cdot \mathcal{M}_{T^S(n)}(\theta) < 1$ for $\theta>0$, the PAVP is upper-bounded by
\begin{align}
  \mathbb{P}\left[\Delta(n)>\zeta\right] \leq \frac{e^{-\theta \zeta} \mathcal{M}_{T^A(n, n+1)}(\theta)}{\left[\mathcal{M}_{T^S(n)}(\theta)\right]^{-1}-\mathcal{M}_{T^A(n, n+1)}(-\theta)}.
\end{align}

\textit{Proof:} Please refer to Appendix A.

\textit{Lemma 2} provides a tractable PAVP upper bound via the MGFs of packet inter-arrival and service times, transforming problem $\mathcal{P}$ into the task of characterizing these MGFs under the constraints of the CF-mMIMO coexistence network. The subsequent sections derive these MGFs to evaluate the bound and optimize the sensing–communication partitioning.

\section{Characterization for Sensing}

This section establishes an analytical framework for the sensing subsystem to characterize the sensory packet arrival process. We first derive the network-wide sensing coverage probability by modeling the collective performance of randomly deployed sensing APs. This result facilitates the derivation of the MGF for the packet arrival process, a requisite input for the PAVP bound in \textit{Lemma 2}.

Sensing APs operate as monostatic radars, distributed according to a PPP $\Phi_s$ with intensity $\lambda_s$. Each sensing AP has a transmit antenna gain $\varsigma_t N$ and a receive antenna gain $\varsigma_r N$, where $\varsigma_t$ and $\varsigma_r$ denote the respective per-antenna gain coefficients. We adopt a sector antenna model with main-lobe beamwidth $2\Theta$. Consequently, only APs within this angular sector contribute to aggregate interference. Propagation follows a bounded path-loss model $\min(1, d^{-\alpha})$, where $d$ is the distance and $\alpha>2$ is the path-loss exponent. Channel gains are subject to i.i.d. Rayleigh fading, denoted by $h \sim \exp(1)$.

\subsection{Network Sensing Coverage Probability}

For CF-mMIMO coexistence networks, the key sensing performance metric is the network sensing coverage probability $\mathbb{P}_{\text{cv}}^{\text{s}}$, defined as the probability that a randomly located target is detected by at least one sensing AP. This metric quantifies the network’s overall situational awareness, rather than the performance of any individual AP.

Using stochastic geometry and Slivnyak's theorem, we analyze the system from the perspective of a typical target at the origin. A sensing AP at $x_i \in \Phi_s$ detects the target if three conditions are met: 1) the target is within the AP's maximum detection range $R$, i.e., $\|x_i\| \le R$;  2) the target lies within the AP's main lobe beamwidth $2\Theta$; and 3) the received echo SINR at this AP exceeds the detection threshold $\delta$. 

We proceed by determining the conditional single-AP detection probability assuming perfect beam alignment, followed by the unconditional network-wide probability accounting for random orientation. Given beam alignment, the received echo power $P_{\text{ech}}$ from a target at range $r = \|x_i\|$ is given by \cite{ref39}:
$$P_{\text{ech}}=\frac{P \varsigma_t \varsigma_r N^2 \lambda_w^2 \sigma}{(4 \pi)^3 r^{2 \alpha}},$$
where $\lambda_w$ is the wavelength. The target's radar cross section (RCS) $\sigma$ follows a Swerling-I model with mean $\bar{\sigma}$ and PDF $p(\sigma)=\bar{\sigma}^{-1} e^{-\sigma/\bar{\sigma}}$ for $\sigma \geq 0$. The sensing SINR $\gamma_s$ is $\gamma_s =\frac{P_{\text{ech}}}{I+N_s},$ where $N_s=k_0 T_0 B_s \chi$ is the sensing noise power ($k_0$ is the Boltzmann's constant, $T_0=290 K$ is the standard temperature, and $\chi=10$ dB is the system loss factor). The aggregate interference $I$ is
$$
I=\sum_{x_i \in \Phi_s} P h_i \min(1, \|x_i\|^{-\alpha}).
$$
This leads to the following theorem, which formally presents the network sensing coverage probability.

\textit{Theorem 1}: In an S\&C coexistence network where sensing APs form a 2D PPP $\Phi_s$ with intensity $\lambda_s$, the network sensing coverage probability $\mathbb{P}_{\text{cv}}^{\text{s}}$ for a target within a maximum range $R$ is
\begin{align}
\mathbb{P}_{\text{cv}}^{\text{s}} = 1 - \exp\left( -2\Theta\lambda_s \int_{0}^{R} r \cdot \mathbb{P}_{\text{sg}}(r) \,dr \right),
\label{sens_prob}
\end{align}
where $\mathbb{P}_{\text{sg}}(r)$ is the single-AP detection probability at distance $r$ conditioned on perfect beam alignment:
\begin{align}
\mathbb{P}_{\text{sg}}(r) &= \exp\left( -\frac{\rho(r)N_s}{P} - 2\Theta\lambda_s \left[ \frac{\rho(r)}{2(1+\rho(r))} \right.\right.\nonumber\\ &\quad\quad\quad\left.\left.+ \frac{\rho(r)}{\alpha\!-\!2} \cdot {}_2F_1\!\left(\!1, 1\!-\!\frac{2}{\alpha}; 2\!-\!\frac{2}{\alpha}; -\rho(r)\!\right) \!\right] \right),
\end{align}
where ${}_2F_1(a,b;c;z)$ is the Gaussian hypergeometric function, and $\rho(r) = \frac{\delta (4\pi)^3r^{2\alpha}}{\varsigma_t \varsigma_r N^2 \lambda_w^2\bar{\sigma}}$.

\textit{Proof:} Please refer to Appendix B.

\textit{Theorem 1} provides a fundamental measure of the network's sensing quality, explicitly linking sensing performance to AP density, antenna characteristics, and propagation parameters.
\vspace{-0.5em}
\subsection{MGF Characterization for Packet Inter-arrival Time}

The sensory packet generation process is coupled to the periodic network scanning outcomes. We model this as a discrete-time process with a scan interval $T_s$. A sensory packet is generated in a scan cycle if network-wide detection is successful, occurring with probability $\mathbb{P}_{\text{cv}}^{\text{s}}$ from \textit{Theorem 1}.

Assuming i.i.d. outcomes across scans, the number of cycles $\iota$ until packet generation follows a geometric distribution with parameter $\mathbb{P}_{\text{cv}}^{\text{s}}$. The inter-arrival time is $T^A(n, n+1) = \iota T_s$, yielding the MGF:
\begin{align}
\mathcal{M}_{T^A(n,n+1)}(\theta) &= \mathbb{E}[e^{\theta \iota  T_s}]\nonumber\\ &= \sum_{\iota =1}^{\infty} e^{\theta \iota  T_s} (1-\mathbb{P}_{\text{cv}}^{\text{s}})^{\iota -1} \mathbb{P}_{\text{cv}}^{\text{s}} \nonumber \\
&= \frac{\mathbb{P}_{\text{cv}}^{\text{s}} e^{\theta T_s}}{1 - (1-\mathbb{P}_{\text{cv}}^{\text{s}})e^{\theta T_s}}\nonumber\\&=\left(\left(e^{-\theta T_s}-1\right){\mathbb{P}_{\text{cv}}^{\text{s}}}^{-1}+1\right)^{-1}.
\label{sens_mgf}
\end{align}
This expression provides the first key component for the PAVP bound in \textit{Lemma 2}, linking the physical-layer sensing performance captured by $\mathbb{P}_{\text{cv}}^{\text{s}}$ to the arrival process statistics.
\vspace{-0.5em}
\section{Uplink Channel Estimation for Communication}
Effective downlink precoding relies on accurate CSI. Leveraging TDD channel reciprocity, the system acquires downlink CSI via uplink pilot training. At the beginning of each coherence interval, a training phase of $\tau_{\mathrm{tr}}$ symbols is dedicated to this purpose. During this phase, users transmit assigned pilot sequences $\psi_k \in \mathbb{C}^{\tau_{\mathrm{tr}} \times 1}$, normalized such that $\|\psi_k\|^2 = 1$. Due to limited orthogonal pilot resources relative to the user density, pilot reuse is inevitable. Users independently select sequences from a set of $\tau_{\mathrm{tr}}$ orthogonal pilots, resulting in pilot contamination whenever $\psi_i^{\mathrm{H}} \psi_k \neq 0$ for interferers $i \neq k$.

The composite signal $\tilde{\mathbf{y}}_{\mathrm{tr}, m} \in \mathbb{C}^{N \times \tau_{\mathrm{tr}}}$ received at AP $m$ is
$$  \tilde{\mathbf{y}}_{\mathrm{tr}, m} = \sum_{i \in \Phi_u} \sqrt{\tau_{\mathrm{tr}} \rho_{\mathrm{tr}}} \mathbf{h}_{m i} \psi_i^{\mathrm{H}} + \mathbf{n}_{\mathrm{tr}, m},$$
where $\rho_{\mathrm{tr}}$ is the normalized pilot SNR, and $\mathbf{n}_{\mathrm{tr}, m}$ is the additive white Gaussian noise matrix with i.i.d. $\mathcal{CN}(0, 1)$ elements. AP $m$ estimates the channel $\mathbf{h}_{mk}$ by projecting the received signal onto $\psi_k$:
\vspace{-0.5em}
\begin{align}
 \tilde{\mathbf{y}}_{m k} &= \frac{1}{\sqrt{\tau_{\mathrm{tr}} \rho_{\mathrm{tr}}}} \tilde{\mathbf{y}}_{\mathrm{tr}, m} \psi_k \nonumber\\&= \mathbf{h}_{m k}+ \sum_{i \in \Phi_u \setminus {k}} \mathbf{h}_{m i} (\psi_i^{\mathrm{H}} \psi_k) + \frac{1}{\sqrt{\tau_{\mathrm{tr}} \rho_{\mathrm{tr}}}} \mathbf{n}_{\mathrm{tr}, m} \psi_k.\nonumber
\end{align}
This processed signal comprises the desired channel response, contamination interference from co-pilot users, and noise.

Assuming known large-scale fading coefficients $\{l_{mi}\}_{i \in \Phi_u}$, AP $m$ computes the linear minimum mean-squared error (LMMSE) estimate estimate $\hat{\mathbf{h}}_{m k}$ based on $\tilde{\mathbf{y}}_{m k}$:
\begin{align}
 \hat{\mathbf{h}}_{m k} \!=\! \mathbb{E}\!\left\{\mathbf{h}_{mk} \tilde{\mathbf{y}}_{m k}^{\mathrm{H}}\right\} \!\left(\mathbb{E}\!\left\{\tilde{\mathbf{y}}_{m k} \tilde{\mathbf{y}}_{m k}^{\mathrm{H}}\right\}\right)^{-1} \!\tilde{\mathbf{y}}_{m k} \!=\! \kappa_{mk} \tilde{\mathbf{y}}_{m k}, \!  
\end{align}
where the scaling coefficient is $\kappa{mk} = l_{mk}/d_{mk}$. Here, $d_{mk}=\sum_{i \in \Phi_u} l_{m i} |\psi_i^{\mathrm{H}} \psi_k|^2 + \frac{1}{\tau_{\mathrm{tr}} \rho_{\mathrm{tr}}}$ represents the total interference-plus-noise power.

Following the derivation of the LMMSE channel estimate $\hat{\mathbf{h}}_{m k}$, we now analyze its statistical properties. By the orthogonality principle of LMMSE estimation, the estimate $\hat{\mathbf{h}}_{m k}$ and the estimation error vector $\tilde{\mathbf{e}}_{m k} = \mathbf{h}_{m k} - \hat{\mathbf{h}}_{m k}$ are uncorrelated, i.e., $\mathbb{E}\{\hat{\mathbf{h}}_{m k} \tilde{\mathbf{e}}_{m k}^{\mathrm{H}}\} = \mathbf{0}$. Since they result from linear operations on Gaussian random variables, they are zero-mean complex Gaussian vectors with distributions $\hat{\mathbf{h}}_{m k} \sim \mathcal{CN}(\mathbf{0}, \sigma_{m k}^2 \mathbf{I}_N)$ and $\tilde{\mathbf{e}}_{m k} \sim \mathcal{CN}(\mathbf{0}, \tilde{\sigma}_{m k}^2 \mathbf{I}_N)$. Their respective variances are
\begin{align}
\sigma_{m k}^2 &= \mathbb{E}\left[ |\hat{h}_{mk,n}|^2 \right] = \kappa_{mk} l_{mk} = \frac{l_{m k}^2}{d_{mk}}, \nonumber\\
\tilde{\sigma}_{m k}^2 &= \mathbb{E}\left[ |\tilde{e}_{mk,n}|^2 \right] = l_{mk} - \sigma_{m k}^2.
\end{align}

To facilitate system-level analysis, we aggregate the channel vectors across all $M$ communication APs to the typical user $k$. Defining the total number of antennas in the cooperative set as $\mathcal{\aleph}=MN$ and assuming independent fading across APs, the aggregated true channel $\mathbf{h}_k=\left[\mathbf{h}_{1 k}^{\mathrm{T}}, \ldots, \mathbf{h}_{M k}^{\mathrm{T}}\right]^{\mathrm{T}}$, its estimate $\hat{\mathbf{h}}_k=\left[\hat{\mathbf{h}}_{1 k}^{\mathrm{T}}, \ldots, \hat{\mathbf{h}}_{M k}^{\mathrm{T}}\right]^{\mathrm{T}}$, and error $\tilde{\mathbf{e}}_k=\left[\tilde{\mathbf{e}}_{1 k}^{\mathrm{T}}, \ldots, \tilde{\mathbf{e}}_{M k}^{\mathrm{T}}\right]^{\mathrm{T}}$ are zero-mean Gaussian vectors. Their block-diagonal covariance matrices are $\mathbf{h}_k \sim \mathcal{CN}(\mathbf{0}, \mathbf{\Lambda}_k)$, $\hat{\mathbf{h}}_k \sim \mathcal{CN}(\mathbf{0}, \mathbf{\Omega}_k)$, and $\tilde{\mathbf{e}}_k \sim \mathcal{CN}(\mathbf{0}, \mathbf{\Lambda}_k - \mathbf{\Omega}_k)$, where
\begin{align}
\mathbf{\Lambda}_k &= \operatorname{diag}\left(l_{1 k} \mathbf{I}_N, \ldots, l_{M k} \mathbf{I}_N\right) \in \mathbb{C}^{\mathcal{\aleph} \times \mathcal{\aleph}},\nonumber \\
\boldsymbol{\Omega}_k &= \operatorname{diag}\left(\sigma_{1 k}^2 \mathbf{I}_N, \ldots, \sigma_{M k}^2 \mathbf{I}_N\right) \in \mathbb{C}^{\mathcal{\aleph} \times \mathcal{\aleph}}.\nonumber
\end{align}

To simplify subsequent expressions, we introduce the auxiliary block diagonal matrix $\mathbf{D} = \operatorname{diag}(d_1 \mathbf{I}_N, \dots, d_M \mathbf{I}_N)$, which allows us to write the relationship $\boldsymbol{\Omega}_k = \mathbf{\Lambda}_k^2 \mathbf{D}^{-1}$. The inverse covariance matrix of the estimate, representing estimation precision, is $\mathbf{C}_k = \mathbf{\Omega}_k^{-1}=\operatorname{diag}(c_{1 k} \mathbf{I}_N, \ldots, c_{M k} \mathbf{I}_N),$
where $c_{m k} = \sigma_{m k}^{-2} = d_m / l_{m k}^2$. 

\vspace{-1em}
\section{Downlink Transmission for Communication}

Following uplink training, communication APs in $\Phi_c$ cooperatively transmit data to the users. This section details the downlink signal model, precoding strategy, and the resulting performance. For a given realization of $M$ communication APs, the signal received by the typical user $k$ is a superposition of transmissions from all cooperative APs and additive noise:
$$
y_{\mathrm{d}, k}=\sqrt{P} \sum_{m=1}^M \mathbf{h}_{m k}^{\mathrm{H}} \mathbf{s}_m+z_{\mathrm{d}, k},
$$
where $\mathbf{h}_{m k} \in \mathbb{C}^{N \times 1}$ is the true channel vector between AP $m$ and user $k$, $\mathbf{s}_m \in \mathbb{C}^{N \times 1}$ is the signal vector from AP $m$, and $z_{\mathrm{d}, k} \sim \mathcal{CN}(0, 1)$ is the normalized AWGN at user $k$. To serve all users concurrently, AP $m$ transmits the signal $\mathbf{s}_m$ that is a linear combination of precoded data symbols:
$$  \mathbf{s}_m=\sqrt{\mu} \sum_{j \in \Phi_u} \mathbf{w}_{m j} q_j,$$
where $q_j$ is the data symbol for user $j$ ($\mathbb{E}\left[|q_j|^2\right]=1$), $\mathbf{w}_{m j}$ is the precoding vector, and $\mu$ is a power normalization factor.

We adopt conjugate beamforming, a low-complexity distributed scheme suitable for CF-mMIMO that relies solely on local CSI. The precoding vector for user \(k\) at AP \(m\) is $
\mathbf{w}_{m k} = c_{m k} \hat{\mathbf{h}}_{m k},$
where \(\hat{\mathbf{h}}_{m k}\) is the local estimate and \(c_{m k}\) is the inverse estimation variance derived in Sec. V. This weighting acts as a statistical channel inversion. To satisfy the average transmit power constraint, $\mu$ is set as
\begin{align} \mu = \left(\mathbb{E}\left[\sum_{j \in \Phi_u} |\mathbf{w}_{mj}|^2\right]\right)^{-1}, \label{p12} \end{align}
where the expectation is taken over all channel realizations.
\vspace{-1em}
\subsection{Downlink SINR Derivation}

Substituting $\mathbf{s}_m$ into $y_{\mathrm{d}, k}$ yields the received signal:
$$  y_{\mathrm{d}, k} = \sqrt{\mu P} \sum_{m \in \Phi_c} \sum_{i \in \Phi_u} \mathbf{h}_{m k}^{\mathrm{H}} (c_{m i} \hat{\mathbf{h}}_{m i}) q_i + z_{\mathrm{d}, k}.$$

We analyze the SINR using the standard "use-and-then-forget" bounding technique. Using aggregate vector notation, $y_{\mathrm{d}, k}$ is decomposed into four distinct components:
$$
\begin{aligned}
y_{\mathrm{d}, k}  =& \underbrace{\sqrt{\mu P} \mathbb{E}\left[\mathbf{h}_k^{\mathrm{H}} \mathbf{C}_k \hat{\mathbf{h}}_k\right] q_k}_{\text{DS}_k: \text{ Desired Signal}} + \underbrace{\sqrt{\mu P} \sum_{i \in \Phi_u \setminus {k}} \mathbb{E}\left[\mathbf{h}_k^{\mathrm{H}} \mathbf{C}_i \hat{\mathbf{h}}_i \right] q_i}_{\text{MUI}_k: \text{ Multi-User Interference}} \\
&+ \underbrace{\sqrt{\mu P} \left( \mathbf{h}_k^{\mathrm{H}} \mathbf{C}_k \hat{\mathbf{h}}_k - \mathbb{E}\left[\mathbf{h}_k^{\mathrm{H}} \mathbf{C}_k \hat{\mathbf{h}}_k\right] \right) q_k}_{\text{BU}_k: \text{ Beamforming Uncertainty}} + \underbrace{z_{\mathrm{d}, k}}_{\text{N}_k: \text{ Noise}},
\end{aligned}
$$
Here, DS$_k$ is the deterministic desired signal component known to the receiver; BU$_k$ is the zero-mean self-interference arising from the receiver's lack of instantaneous channel knowledge; MUI$_k$ is the zero-mean interference from other users due to pilot contamination; and N$_k$ is the noise.

Treating the uncertainty and interference terms as uncorrelated noise, the effective SINR $\gamma_k$ for user $k$ is: 
\begin{align}
\gamma_k= \frac{\left|\mathbb{E}\left[\mathbf{h}_k^H \mathbf{C}_k \hat{\mathbf{h}}_k\right]\right|^2}{\operatorname{var}\left[\mathbf{h}_k^H \mathbf{C}_k \hat{\mathbf{h}}_k\right]\!+\!\sum_{i \in \Phi_u \!\setminus {k}}\! \mathbb{E}\left[\left|\mathbf{h}_k^H \mathbf{C}_i \hat{\mathbf{h}}_i\right|^2\right]\!+\!\frac{1}{\mu P}}.  
\label{p15}
\end{align}

Eq. \eqref{p15} provides the SINR for a given network realization, conditioned on the large-scale fading. However, its dependence on the random AP locations makes direct network-wide analysis intractable, motivating the use of deterministic equivalents.

\vspace{-1em}
\subsection{Deterministic Equivalent SINR Analysis}

The SINR $\gamma_k$ in Eq. \eqref{p15} depends on small-scale fading and random AP locations, making direct analysis intractable. We employ deterministic equivalent (DE) analysis \cite{ref32} to derive an asymptotic approximation that becomes exact as the system dimensions grow large. Formally, a DE for a sequence of random variables $\{X_n\}_{n=1}^\infty$ is a deterministic sequence $\{\bar{X}_n\}_{n=1}^\infty$ such that $X_n - \bar{X}_n \xrightarrow{a.s.} 0$ as $n \rightarrow \infty$. Applying DE techniques to Eq. \eqref{p15} yields the following lemma for the large-scale spatial limit.

\textit{Lemma 3: }Conditioned on a fixed spatial realization of APs $\Phi_c$ and users $\Phi_u$, the downlink DE SINR for user $k$ is
\vspace{-0.7em}
\begin{align}
 \bar{\gamma}_k \asymp \frac{\mathcal{\aleph}}{\frac{1}{\mathcal{\aleph}} \sum_{i \in \Phi_u} \operatorname{tr} \mathbf{D \Lambda}_i^{-2}\left(\mathbf{\Lambda}_k+\frac{\mathcal{\aleph}}{P}\right)-1},   
 \label{p17}
\end{align}
where $\asymp$ denotes the deterministic equivalent. 


\textit{Proof:} Please refer to Appendix C.

\textit{Lemma 3} provides a deterministic proxy for SINR that averages out small-scale fading, formally capturing the channel hardening phenomenon inherent to CF-mMIMO. As the number of cooperative antennas $\aleph$ increases, the SINR fluctuations vanish, converging to $\bar{\gamma}_k$. This convergence allows us to treat the SINR as constant over coherence blocks sharing the same large-scale fading realization, significantly simplifying the subsequent service time analysis.


\vspace{-1em}
\subsection{Network Communication Coverage Probability}
While the DE analysis eliminates the randomness from small-scale fading, the asymptotic SINR $\bar{\gamma}_k$ in \textit{Lemma 3} remains stochastic due to the random topology of $\Phi_c$ and $\Phi_u$. We employ stochastic geometry to characterize this network-level randomness via the communication coverage probability, defined as the complementary cumulative distribution function (CCDF) of the downlink SINR.

\textit{Definition 2:} A typical user is considered to be in communication coverage if its downlink SINR $\gamma_k$ exceeds the threshold $\gamma_{\mathrm{th}}$. The communication coverage probability is thus given by
\vspace{-0.5em}
$$
\mathbb{P}_{\mathrm{cv}}^{\mathrm{c}}(\gamma_{\mathrm{th}}) = \mathbb{P}[\gamma_k > \gamma_{\mathrm{th}}].
$$

The following theorem establishes a tractable lower bound for this metric.

\textit{Theorem 2:} In a pilot-contaminated CF-mMIMO network with independent homogeneous PPPs for APs and users, the communication coverage probability is lower bounded by
\vspace{-0.7em}
\begin{align} 
\mathbb{P}_{\text{cv}}^{\text{c}}(\gamma_\mathrm{th}) \geq 1-\left(1-e^{-\eta_c \gamma_\mathrm{th}\psi(\tilde{\mathcal{\aleph}})}\right)^{\tilde{\mathcal{\aleph}}}, \label{eq:20}
\end{align}
where $\tilde{\mathcal{\aleph}}=\mathbb{E}[\mathcal{\aleph}]$, $\eta_c=\tilde{\mathcal{\aleph}}(\tilde{\mathcal{\aleph}}!)^{-\frac{1}{\tilde{\mathcal{\aleph}}}}$ and
\vspace{-0.7em}
\begin{align}
\psi(\tilde{\mathcal{\aleph}}) = \frac{\lambda_u}{\lambda_c \tau_{tr} } \left( \lambda_u \!+\! \frac{\alpha-2}{\alpha\pi  \rho_{tr} } \right) \left( 1 \!+ \!\frac{(\alpha-1)\tilde{\mathcal{\aleph}}}{(\alpha-2)P} \right) - 1.
\label{psi}
\end{align}

\textit{Proof:} Please refer to Appendix D.

This lower bound reveals that coverage $\mathbb{P}_{\text{cv}}^{\text{c}}$ degrades with increasing user intensity $\lambda_u$ and pilot contamination (inverse to $\tau_{tr}$), both of which heighten aggregate multi-user interference. Crucially, performance is primarily governed by the ratio of user density to AP density $\lambda_u / \lambda_c$ rather than absolute numbers. As $\lambda_c$ increases relative to $\lambda_u$, coverage improves and eventually saturates, reflecting the transition from a noise-limited to an interference-limited regime.

\vspace{-1em}
\subsection{MGF Characterization for Communication}

We utilize the communication coverage probability from \textit{Theorem 2} to derive the MGF of the packet service time. This MGF statistically describes the time required to successfully deliver a sensory packet and serves as the final component for the network-wide PAVP analysis within our SNC framework.

As defined in Sec. II-A, the service process is a discrete-time system with intervals of $T_c \!=\! \tau_c / B_c$. Accounting for uplink training, the effective data transmission duration is $\!\tau_d \!\!=\! \tau_c \!- \!\tau_{\mathrm{tr}}\!$ symbols. A packet transmission follows a retransmission protocol where the number of attempts $J$ for successful delivery follows a geometric distribution with DEP $\epsilon$. The DEP depends on the instantaneous SINR $\gamma$, as dictated by FBC theory. For a packet of $L$ bits transmitted over $\tau_d$ symbols, the effective rate is $R_{eff} = L / (\tau_d / B_c)$. Solving Eq. \eqref{a} for $\epsilon$ yields
$$
\epsilon(\gamma) = Q\left( \frac{\ln(1+\gamma) - \frac{L \ln 2}{\tau_d}}{\sqrt{V(\gamma)/\tau_d}} \right).
$$
Conditioned on a fixed SINR $\gamma$, the MGF of the service time $T^S(n) = J \cdot T_c$ is derived from the geometric distribution:
\begin{align}
\mathcal{M}_{T^S(n)|\Gamma}(\theta|\gamma) &= \mathbb{E}\left[e^{\theta T^S(n)} | \Gamma=\gamma\right] \nonumber\\&= \mathbb{E}\left[e^{\theta \cdot J \cdot T_c}\right] \nonumber\\
&= \sum_{j=1}^{\infty} e^{\theta j T_c} \epsilon(\gamma)^{j-1}(1-\epsilon(\gamma)) \nonumber\\
&= \frac{(1-\epsilon(\gamma)) e^{\theta T_c}}{1 - \epsilon(\gamma) e^{\theta T_c}}.\nonumber
\end{align}
Transmission is viable only when the user is in coverage ($\Gamma \ge \gamma_\mathrm{th}$). To obtain the unconditional MGF, we average over the SINR distribution conditioned on this event. Utilizing the tractable lower bound from \textit{Theorem 2} ensures a worst-case performance guarantee. The corresponding conditional PDF is
\begin{align}
f_{\Gamma|\Gamma \ge \gamma_\mathrm{th}}^{\text{bound}}(\gamma) &= \frac{f_{\Gamma}^{\text{bound}}(\gamma)}{P(\Gamma \ge \gamma_\mathrm{th})} \nonumber\\&= \frac{\tilde{\mathcal{\aleph}} \eta_c \psi(\tilde{\mathcal{\aleph}}) e^{-\eta_c \gamma \psi(\tilde{\mathcal{\aleph}})} \left(1-e^{-\eta_c \gamma \psi(\tilde{\mathcal{\aleph}})}\right)^{\tilde{\mathcal{\aleph}}-1}}{1-\left(1-e^{-\eta_c \gamma_\mathrm{th} \psi(\tilde{\mathcal{\aleph}})}\right)^{\tilde{\mathcal{\aleph}}}}.\nonumber
\end{align}
Applying the law of total expectation, the MGF of the packet service time, conditioned on the user being in coverage, is
\begin{align} 
\mathcal{M}_{T^S(n)}(\theta) &= \mathbb{E}\left[e^{T^S(n) \theta } | \Gamma \ge \gamma_\mathrm{th}\right] \nonumber\\
&= \int_{\gamma_\mathrm{th}}^{\infty} \mathcal{M}_{T^S(n)|\Gamma}(\theta|\gamma) f_{\Gamma|\Gamma \ge \gamma_\mathrm{th}}^{\text{bound}}(\gamma) \, \mathrm{d}\gamma \nonumber\\
&= \frac{\tilde{\mathcal{\aleph}} \eta_c \psi(\tilde{\mathcal{\aleph}})}{\mathbb{P}_{\text{cv}}^{\text{c}}(\gamma_\mathrm{th})} \!\!\int_{\gamma_\mathrm{th}}^{\infty}\! \frac{(1\!-\!\epsilon(\gamma)) e^{-\eta_c \gamma \psi(\tilde{\mathcal{\aleph}})}}{e^{-\theta T_c} - \epsilon(\gamma)}\!  \nonumber\\
&\quad\quad\quad\quad\quad\quad\quad\!\cdot\left(1\!-\!e^{-\eta_c \gamma \psi(\tilde{\mathcal{\aleph}})}\right)^{\tilde{\mathcal{\aleph}}\!-\!1} \!\mathrm{d}\gamma.\!
\label{eq:mgf_comm}
\end{align}

\textit{Remark 1:} Eq. \eqref{eq:mgf_comm} constitutes a valid upper bound on the true service time MGF. This is methodologically essential for the SNC framework, as using a service time upper bound (derived from a coverage lower bound) ensures a conservative and reliable worst-case guarantee for the final PAVP analysis. This expression encapsulates the combined effects of stochastic AP deployment, imperfect CSI, and FBC retransmissions.

\begin{figure*}[htbp]
\vspace{-1em}
\begin{align}
\Upsilon= \frac{e^{-\theta \zeta} \left(\left(e^{-\theta T_s}-1\right){\mathbb{P}_{\text{cv}}^{\text{s}}}^{-1}+1\right)^{-1}}{ \left( \frac{\tilde{\mathcal{\aleph}} \eta_c \psi(\tilde{\mathcal{\aleph}})}{\mathbb{P}_{\text{cv}}^{\text{c}}(\gamma_\mathrm{th})} \displaystyle \int_{\gamma_\mathrm{th}}^{\infty} \frac{(1-\epsilon(\gamma)) e^{-\eta_c \gamma \psi(\tilde{\mathcal{\aleph}})}}{e^{-\theta T_c} - \epsilon(\gamma)} \left(1-e^{-\eta_c \gamma \psi(\tilde{\mathcal{\aleph}})}\right)^{\tilde{\mathcal{\aleph}}-1} \mathrm{d}\gamma \right)^{-1} \!- \left(\left(e^{\theta T_s}-1\right){\mathbb{P}_{\text{cv}}^{\text{s}}}^{-1}+1\right)^{-1} }.
\label{eq:final_bound}
\end{align}
{\noindent}\rule[0pt]{18cm}{0.05em}
\vspace{-1em}
\end{figure*}
\vspace{-0.5em}
\section{PAVP Analysis and Optimization of Communication and Sensing}
We now synthesize the derived MGFs for the arrival and service processes to establish a tractable upper bound on the network-wide PAVP within the SNC framework. Subsequently, we minimize this bound to determine the optimal functional partitioning of APs between sensing and communication.
\vspace{-0.5em}
\subsection{PAVP Upper Bound}

We begin by considering the PAVP for a user in communication coverage. The MGF of the packet inter-arrival time $\mathcal{M}_{T^A}(\theta)$ (Eq. \eqref{sens_mgf}) captures the statistics of radar-driven packet generation. Concurrently, the service time MGF $\mathcal{M}_{T^S}(\theta)$ (Eq. \eqref{eq:mgf_comm}) accounts for the CF-mMIMO communication dynamics, including pilot contamination and ARQ retransmissions. Substituting these into the general SNC bound from \textit{Lemma 2} yields the conditional PAVP upper bound $\Upsilon$ given in Eq. \eqref{eq:final_bound} at the top of page, valid for users in communication coverage.

To establish a network-wide metric, we further account for users outside communication coverage. For such users, packet delivery is unavailable, causing the PAoI to grow unbounded and inevitably violate the threshold $\zeta$. Thus, the PAVP for an out-of-coverage user is effectively 1. The probability that a user is both in coverage and satisfies the PAoI constraint is lower-bounded by $\mathbb{P}_{\text{cv}}^{\text{c}} \cdot (1-\Upsilon)$. Combining both in-coverage and out-of-coverage cases yields the following theorem.

\textit{Theorem 3:} The network-wide statistical PAVP for a user in the CF-mMIMO coexistence network is upper-bounded by
\vspace{-0.5em}
\begin{align}
 \Upsilon_{\text{nw}} = 1 - \mathbb{P}_{\text{cv}}^{\text{c}} \cdot (1-\Upsilon),
\label{eq:unconditional_pavp}
\end{align}
where $\mathbb{P}_{\text{cv}}^{\text{c}}$ is the communication coverage probability lower bound from Eq. \eqref{eq:20} and $\Upsilon$ is the conditional PAVP upper bound from Eq. \eqref{eq:final_bound}.
\vspace{-0.5em}
\subsection{PAVP Upper Bound Optimization}
The network-wide PAVP upper bound $\Upsilon_{\text{nw}}$ encapsulates the network-level S\&C trade-off. Both the sensing performance (via $\mathbb{P}_{\text{cv}}^{\text{s}}$ in $\mathcal{M}_{T^A}$) and communication performance (via $\mathbb{P}_{\text{cv}}^{\text{c}}$ and $\mathcal{M}_{T^S}$) depend on the partitioning factor $\beta$. We formulate the PAVP minimization problem as
\vspace{-0.5em}
\begin{align}
    \tilde{\mathcal{P}}: \quad &\min_{0 \leq \beta \leq 1} \quad \Upsilon_{\text{nw}}(\beta)  \nonumber \\
    &\text{s.t.} \quad \mathcal{M}_{T^A(n, n+1)}(-\theta) \cdot \mathcal{M}_{T^S(n)}(\theta) < 1, \nonumber
\end{align}
where the constraint ensures the stability of the underlying FCFS queuing system, as required by \textit{Lemma 2}.

The objective function is analytically involved, and a closed-form optimizer is difficult to obtain. Nevertheless, its boundary behavior offers useful intuition. 
\begin{enumerate}
\item Sensing-Only Deployment ($\beta \to 0$): As the communication AP density $\lambda_c \to 0$, the communication coverage $\mathbb{P}_{\text{cv}}^{\text{c}}$ vanishes. Eq. \eqref{eq:unconditional_pavp} implies $\Upsilon_{\text{nw}} \to 1$. Physically, while sensory packets are generated frequently, the lack of a transport network prevents delivery, resulting in certain PAoI violations.
\item Communication-Only Deployment ($\beta \to 1$): As sensing AP density $\lambda_s \to 0$, the sensing coverage vanishes, leading to infinite packet inter-arrival times. Consequently, $\Upsilon_{\text{nw}} \to 1$ due to extreme data staleness. Physically, networks without sufficient sensing capability cannot generate timely information, rendering the communication subsystem ineffective in maintaining freshness. 
\end{enumerate}

Therefore, under a fixed shared resource budget, neither extreme allocation can sustain network-wide freshness. We emphasize, however, that the main value of the proposed framework is not merely in asserting the existence of an interior minimizer. Rather, it lies in providing a tractable optimization surrogate that quantifies how the minimizing partition factor shifts with sensing quality, communication density, user load, pilot overhead, and PAoI requirements. Although the derived bound is conservative in absolute PAVP values, the numerical results show that the optimizer of the bound is highly consistent with the optimizer observed in Monte Carlo simulations. Hence, $\tilde{\mathcal{P}}$ offers an efficient low-complexity means of identifying near-optimal coexistence configurations without exhaustive simulation-based search.

Although a closed-form solution is intractable, the optimal partitioning factor can be efficiently obtained through a one-dimensional line search. This result provides a practical guideline for network planning and slow-timescale coexistence adaptation in large-scale CF-mMIMO systems serving xURLLC applications.





\begin{table}[htbp]
    \centering
    \caption{Default Parameter Settings}
    \label{tab:sim_params}
    
    \setlength{\tabcolsep}{4pt} 
    \footnotesize 
    
    \begin{tabular}{cccccc}
    \hline
    \textbf{Param.} & \textbf{Value} & \textbf{Param.} & \textbf{Value} & \textbf{Param.} & \textbf{Value} \\
    \hline
    $\lambda$ & 100 $\mathrm{AP/km^2}$ & $\lambda_u$ & 30 $\mathrm{UE/km^2}$ & $L$ & 100 bits \\
    $T_s$ & 1 ms & $\zeta$ & 5 ms & $B_s, B_c$ & 1 MHz \\
    $N$ & 10 & $\alpha$ & 2.1 & $P$ & 1W \\
    $\beta$ & 0.3 & $\varsigma_t, \varsigma_r$ & 20 dBi & $\bar{\sigma}$ & 20 dBsm \\
    $R$ & 500 m & $\delta$ & -10 dB & $\Theta$ & $\pi$ rad \\
    $N_s$ & -104 dBm & $\tau_c$ & 210 Sym. & $\tau_{tr}$ & 10 Sym. \\
    $\mathcal{N}$ & 200 Sym. & $\rho_{tr}$ & 23 dBm & $\gamma_\mathrm{th}$ & 0 dB \\
    \hline
    \end{tabular}

    \vspace{1.5em} 

    \includegraphics[width=0.85\columnwidth]{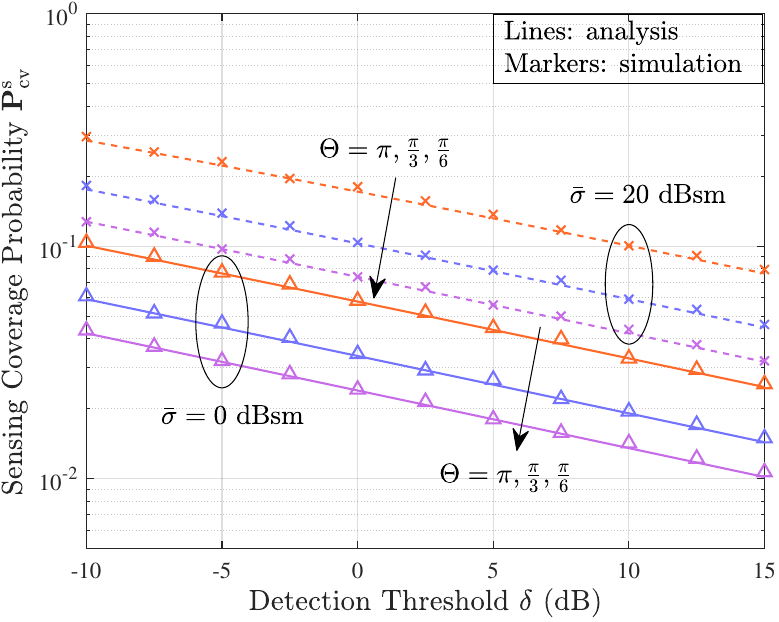}
    
    \captionsetup{justification=raggedright,singlelinecheck=false}
    \captionof{figure}{\small $\mathbb{P}_{\text{cv}}^{\text{s}}$ vs. $\delta$ for various $\bar{\sigma}$ and $\Theta$.}
    \label{fig:sensing_coverage}
    
    \vspace{-1.5em} 
\end{table}

\vspace{-0.5em}
\section{Numerical Results}
This section presents numerical results to validate our analytical framework and extract design insights for S\&C coexistence systems targeting xURLLC applications. Default parameter settings are listed in TABLE \ref{tab:sim_params}. We utilize Monte Carlo simulations to corroborate our analytical findings. Each simulation point represents an average over $10^7$ independent spatial realizations of the PPP network topology. For figures involving queuing dynamics (Figs. 6-8), we simulate $10^7$ packets per spatial realization to ensure statistical accuracy.




Fig. \ref{fig:sensing_coverage} investigates the network sensing coverage probability $\mathbb{P}_{\text{cv}}^{\text{s}}$ against the detection threshold $\delta$, considering varying radar beamwidths $\Theta$ and average target RCS $\bar{\sigma}$. The tight agreement between our analytical model from \textit{Theorem 1} and the Monte Carlo simulations confirms its accuracy. As expected, coverage degrades with more stringent thresholds $\delta$ but improves substantially for targets with larger RCS due to stronger echo power. Notably, wider beams provide superior coverage compared to narrower beams. In a network with random AP orientations, a wider beam significantly increases the target illumination probability. This spatial coverage gain outweighs the increased susceptibility to co-channel interference, establishing a crucial design principle: for wide-area surveillance with unknown target locations, employing broad-beam APs improves detection efficacy.

\begin{figure*}[htbp]
\vspace{-2em}
	\begin{minipage}{0.33\linewidth}
    \small
		\centering
		\includegraphics[width=\linewidth]{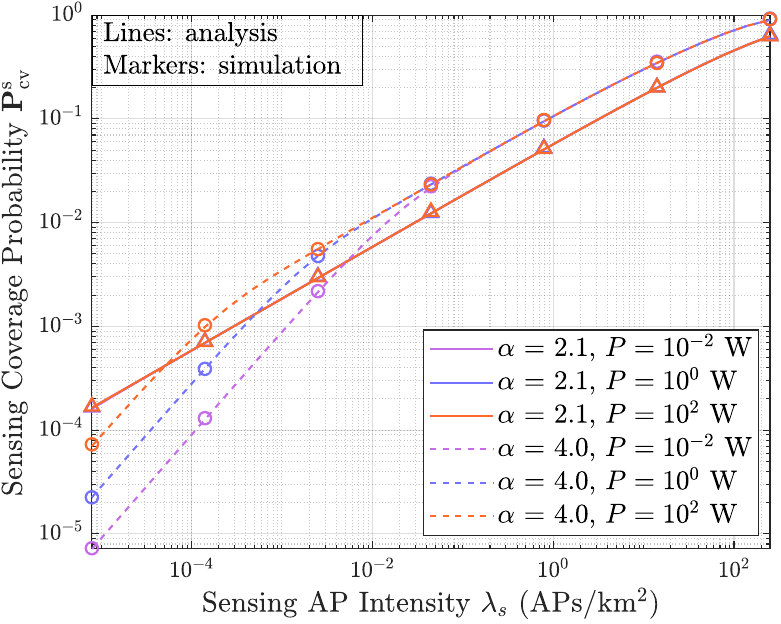}
        \captionsetup{justification=raggedright,singlelinecheck=false} 
		\caption{\small $\mathbb{P}_{\text{cv}}^{\text{s}}$ vs. $\lambda_s$ for various $P$ and $\alpha$.}
    	\label{fig:sensing_vs_lambda_s}
	\end{minipage}
	\begin{minipage}{0.33\linewidth}
    \small
		\centering
		\includegraphics[width=\linewidth]{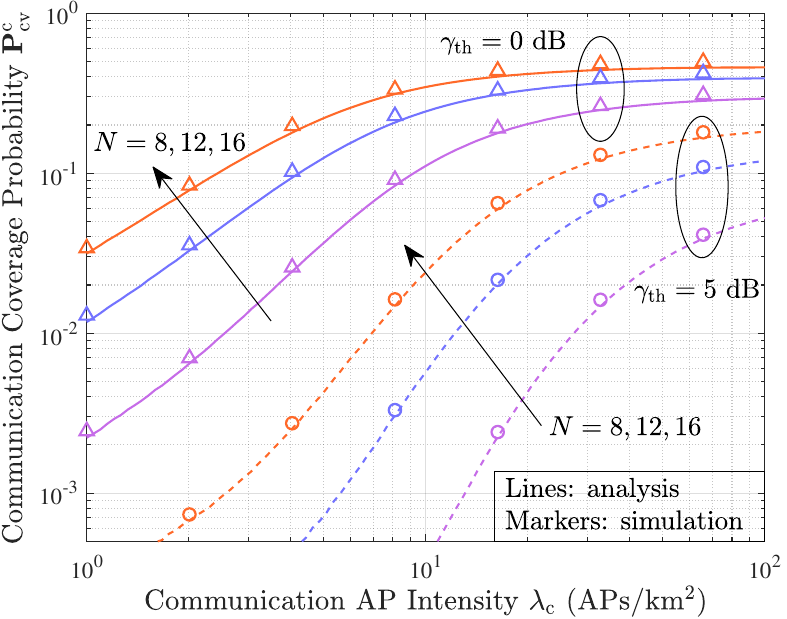}
        \captionsetup{justification=raggedright,singlelinecheck=false} 
		\caption{\small $\mathbb{P}_{\text{cv}}^{\text{c}}$ vs. $\lambda_c$ for various $N$ and $\gamma_\mathrm{th}$.}
		\label{fig:comm_coverage_1}
	\end{minipage}
	\begin{minipage}{0.33\linewidth}
    \small
		\centering
		\includegraphics[width=\linewidth]{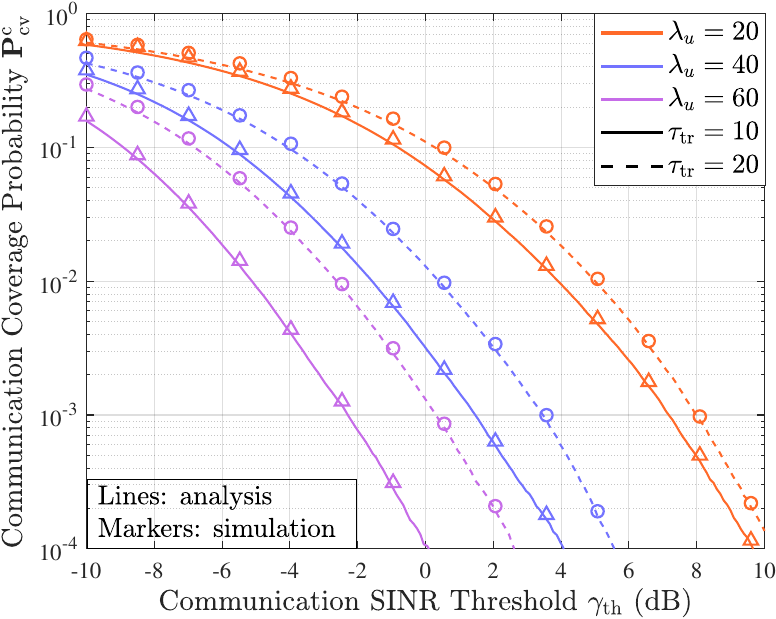}
        \captionsetup{justification=raggedright,singlelinecheck=false} 
		\caption{\small $\mathbb{P}_{\text{cv}}^{\text{c}}$ vs. $\gamma_{\text{th}}$ for various $\lambda_u$ and $\tau_{\text{tr}}$.}
		\label{fig:comm_coverage}
	\end{minipage}
\end{figure*}

Fig. \ref{fig:sensing_vs_lambda_s} examines the impact of sensing AP intensity $\lambda_s$ on coverage under different transmit powers and path-loss exponents. The analytical curves align closely with simulations. Coverage improves monotonically with $\lambda_s$ as the likelihood of AP proximity to the target increases. The results reveal a critical interplay between the path-loss exponent and AP density. In sparse, noise-limited regimes, a lower exponent is preferable due to better signal propagation. Conversely, in dense, interference-limited regimes, a higher exponent becomes advantageous because it attenuates aggregate interference more effectively than the desired signal. Furthermore, increasing transmit power offers negligible gains in dense networks. This suggests that in high-path-loss urban environments, network densification is a far more effective strategy for enhancing sensing performance than power boosting.

Fig. \ref{fig:comm_coverage_1} depicts the communication coverage probability $\mathbb{P}_{\text{cv}}^{\text{c}}$ versus AP intensity $\lambda_c$, validating the tightness of the lower bound in \textit{Theorem 2}. Coverage improves with higher AP density (enhanced macro-diversity) and larger antenna arrays (greater array gain). However, densification exhibits diminishing returns as the curves saturate at high $\lambda_c$, marking the transition from a deployment-limited to an interference-limited regime. This provides a key architectural insight: while initial densification is beneficial, leveraging larger antenna arrays becomes more resource-efficient than further densification in highly dense networks.

\begin{figure*}[htbp]
\begin{minipage}{0.33\linewidth}
    \small
		\centering
		\includegraphics[width=\linewidth]{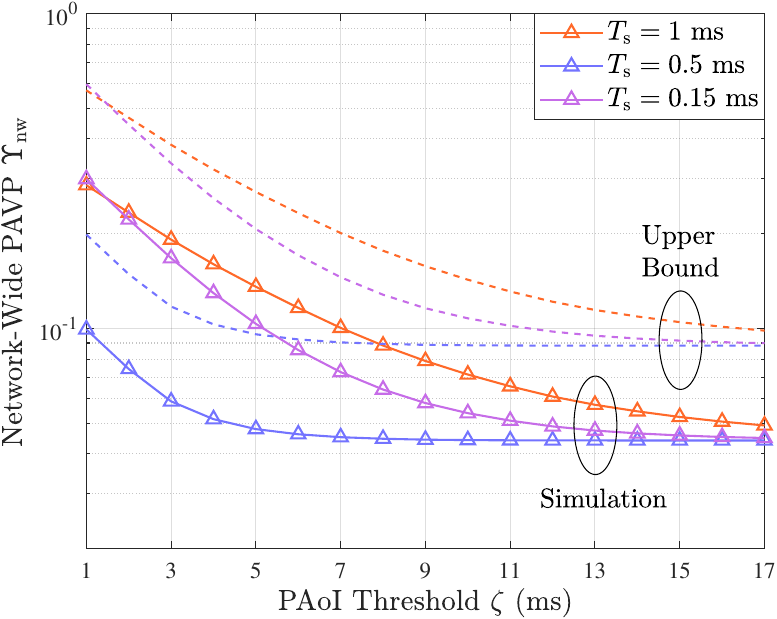}
        \captionsetup{justification=raggedright,singlelinecheck=false} 
		\caption{\small $\Upsilon_{\text{nw}}$ vs. $\zeta$ for various $T_s$.}
		\label{th3fig1}
	\end{minipage}
    \begin{minipage}{0.33\linewidth}
    \small
		\centering
		\includegraphics[width=\linewidth]{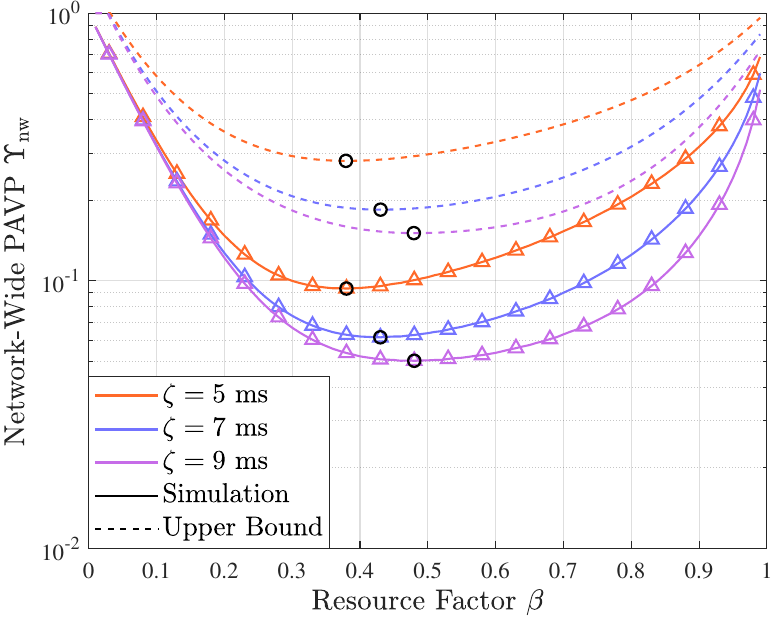}
        \captionsetup{justification=raggedright,singlelinecheck=false} 
		\caption{\small $\Upsilon_{\text{nw}}$ vs. $\beta$ for various $\zeta$.}
		\label{th3fig2}
	\end{minipage}
	\begin{minipage}{0.33\linewidth}
    \small
		\centering
		\includegraphics[width=\linewidth]{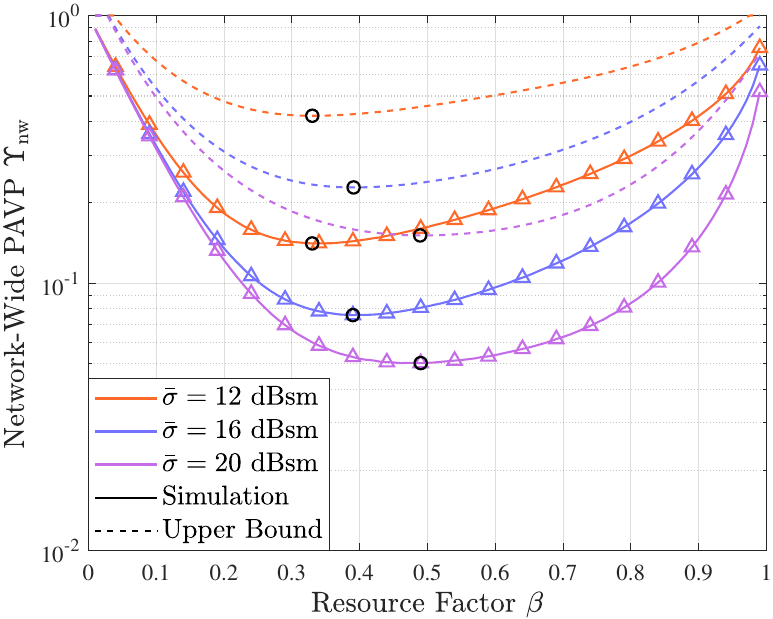}
        \captionsetup{justification=raggedright,singlelinecheck=false} 
		\caption{\small $\Upsilon_{\text{nw}}$ vs. $\beta$ for various $\bar{\sigma}$.}
    	\label{th3fig3}
	\end{minipage}
	\vspace{-1em}
\end{figure*}

Fig. \ref{fig:comm_coverage} explores communication coverage versus SINR threshold $\gamma_{\text{th}}$. Increasing user intensity $\lambda_u$ severely degrades coverage due to heightened pilot contamination and collision probability. In contrast, increasing the pilot sequence length from $\tau_{\text{tr}}=10$ to $\tau_{\text{tr}}=20$ significantly mitigates this by enhancing orthogonality. This highlights a primary bottleneck in large-scale cell-free systems: in dense user scenarios, allocating more coherence resources to uplink training is essential for communication reliability, even at the cost of reduced data transmission time.

Fig. \ref{th3fig1} plots the network-wide PAVP against the PAoI threshold $\zeta$ for various sensing intervals $T_s$. The figure reveals a non-monotonic relationship between PAVP and $T_s$. An intermediate value ($T_s=0.5$ ms) consistently outperforms both rapid ($T_s=0.15$ ms) and slow ($T_s=1$ ms) sensing. Rapid sensing reduces the generation interval but risks overwhelming the communication queue, while slow sensing avoids congestion but introduces staleness. The crossover between the $T_s=1$ ms and $T_s=0.15$ ms curves underscores this trade-off: for stringent deadlines ($\zeta < 2$ ms), the queuing delay is the dominant bottleneck, favoring slower sensing. For relaxed deadlines ($\zeta > 2$ ms), the generation interval dominates, favoring rapid sensing. Thus, the optimal sensing period must balance the update rate with the network's delivery capacity.

Fig. \ref{th3fig2} depicts the network-wide PAVP upper bound $\Upsilon_{\text{nw}}$ against the resource allocation factor $\beta$. The black circles mark the optimal $\beta^*$ values that minimize PAVP. The curves, comparing our analytical bound with simulations across varying PAoI thresholds $\zeta$, exhibit a distinct quasi-convex shape. This empirically confirms the network-level S\&C trade-off: extreme allocations (i.e., $\beta \to 0$ or $\beta \to 1$) result in high PAVP, validating the existence of an optimal allocation $\beta^*$ that minimizes information staleness. Crucially, $\beta^*$ depends on the application's AoI requirement. As the deadline becomes more stringent, the optimal $\beta^*$ shifts leftward. This reveals a critical design principle: for services with ultra-tight PAoI constraints, the system must prioritize rapid packet generation by dedicating more resources to sensing; conversely, for services with more relaxed deadlines, it is more effective to strengthen the communication infrastructure to ensure reliable delivery.

Fig. \ref{th3fig3} presents the network-wide PAVP upper bound $\Upsilon_{\text{nw}}$ against $\beta$ for different average target RCS $\bar{\sigma}$. The PAVP improves significantly as the target's average RCS increases, since a larger RCS enhances the sensing coverage probability, facilitating more frequent and timely packet generation. Importantly, the figure reveals that $\beta^*$ shifts rightward as sensing conditions improve. This is because a higher RCS makes the target easier to detect, allowing the network to achieve the requisite sensing performance with fewer sensing APs. Consequently, a larger portion of APs can be allocated to communication to enhance delivery reliability and reduce queuing delays, thereby further lowering the overall PAVP. This highlights that the optimal S\&C resource partition must adapt to the operational environment. Specifically, in scenarios with favorable sensing conditions, the performance bottleneck shifts towards communication, warranting a greater allocation of resources to the communication infrastructure to maximize network-wide information freshness.

It is worth emphasizing that the main value of the proposed bound lies not only in providing a conservative performance guarantee, but also in accurately tracking the optimization trend and the minimizing partition factor with much lower complexity than exhaustive simulation-based search.


\vspace{-1em}
\section{Conclusions}
In this paper, we developed a tractable analytical framework for network-wide PAoI guarantee in CF-mMIMO networks with S\&C coexistence toward xURLLC. By combining stochastic geometry and stochastic network calculus, we characterized the sensing-driven packet arrival process and the communication-limited service process under stochastic topology, imperfect CSI, and finite blocklength transmission, and thereby derived a conservative upper bound on the network-wide PAVP. The resulting analysis reveals how spatial AP partitioning shapes the coexistence trade-off between update generation and reliable packet delivery in large-scale distributed networks. The proposed bound accurately tracks the performance trend and yields a minimizing partition factor that closely matches the simulation optimum. This makes the framework a useful low-complexity tool for coexistence-oriented network design and efficient PAoI guarantee in large-scale CF-mMIMO systems.


%




\ifCLASSOPTIONcaptionsoff
  \newpage
\fi



%


\flushbottom

\vspace{-0.5em}
\appendices
\vspace{-0.5em}
\section{Proof of Lemma 2}
Using the equality derived for $\Delta(n)$ in Eq. \eqref{d} and applying the union bound, the MGF of the PAoI can be bounded by
\vspace{-0.5em}
\begin{align}
&\mathcal{M}_{\Delta(n)}(\theta)=\mathbb{E}\left[e^{\Delta(n)\theta}\right] \nonumber\\
\leq &\sum_{v=1}^{n+1} \mathbb{E}\left[e^{\theta T^S(v, n+1)}\right] \mathbb{E}\left[e^{-\theta T^A(v, n+1)}\right]\mathbb{E}\left[e^{\theta T^A(n, n+1)}\right] \nonumber,
\end{align}

Assuming that the inter-arrival times and service times are i.i.d., we have 
\vspace{-0.5em}
$$\mathbb{E}\left[e^{-\theta T^A(v, n+1)}\right] = \left(\mathcal{M}_{T^A(n, n+1)}(-\theta)\right)^{n-v+1},$$
and 
$$\mathbb{E}\left[e^{\theta T^S(v, n+1)}\right] = \left(\mathcal{M}_{T^S(n)}(\theta)\right)^{n-v+2}.$$ 
Substituting these back into the MGF bound yields
\vspace{-0.5em}
\begin{align}
&\mathcal{M}_{\Delta(n)}(\theta) \nonumber\\
\leq&\sum_{v=1}^{n+1} \mathbb{E}\left[e^{\theta T^S(v, n+1)}\right] \mathbb{E}\left[e^{-\theta T^A(v, n+1)}\right]\mathbb{E}\left[e^{\theta T^A(n, n+1)}\right] \nonumber\\
=&\mathbb{E}\left[e^{\theta T^A(n, n+1)}\right]\mathbb{E}\left[e^{\theta T^{\mathrm{s}}(n)}\right] \nonumber\\
&\quad \quad \quad \quad \quad \cdot \sum_{v=1}^{n+1}\left(\mathbb{E}\left[e^{\theta T^{\mathrm{s}}(n)}\right]\mathbb{E}\left[e^{-\theta T^{A}(n, n+1)}\right]\right)^{(n-v+1)} \nonumber \\
\leq &\mathbb{E}\!\left[e^{\theta T^A(n, n\!+\!1)}\!\right]\!\mathbb{E}\!\left[e^{\theta T^{\mathrm{s}}(n)}\!\right] \!\sum_{v=0}^{\infty}\!\left(\!\mathbb{E}\!\left[e^{\theta T^{\mathrm{s}}(n)}\!\right]\!\mathbb{E}\!\left[e^{-\theta T^{A}(n, n\!+\!1)}\!\right]\!\right)^{\!v}.\nonumber
\end{align}
Under the stability condition, this geometric series converges as $n \to \infty$, giving
\vspace{-0.5em}
$$
\mathcal{M}_{\Delta(n)}(\theta) \leq \frac{\mathcal{M}_{T^A(n, n+1)}(\theta) \mathcal{M}_{T^S(n)}(\theta)}{1-\mathcal{M}_{T^A(n, n+1)}(-\theta) \mathcal{M}_{T^S(n)}(\theta)}.
$$
Applying the Chernoff bound $\mathbb{P}[X \geq x] \leq e^{-\theta x} \mathcal{M}_X(\theta)$ to the MGF bound of PAoI gives the result in \textit{Lemma 2}. \hspace*{\fill}\qed
\vspace{-0.5em}
\section{Proof of Theorem 1}
We first derive the network sensing outage probability $\mathbb{P}_{\text{out,s}}$, defined as the probability that a typical target at the origin remains undetected by any sensing AP. The coverage probability is subsequently $\mathbb{P}_{\text{cv}}^{\text{s}} = 1 - \mathbb{P}_{\text{out,s}}$.

Let $p(\mathbf{x})$ be the unconditional detection probability for an AP at $\mathbf{x} \in \mathbb{R}^2$. This event necessitates that the AP is within range $R$, the antenna is aligned with the target, and the SINR threshold is met. We can express this as
\vspace{-0.5em}
$$
p(\mathbf{x}) = \frac{\Theta}{\pi} \cdot \mathbb{P}_{\text{sg}}(\|\mathbf{x}\|) \cdot \mathbb{I}(\|\mathbf{x}\| \le R),
$$
where $\mathbb{P}_{\text{sg}}(r)$ is the conditional detection probability at distance $r=\|\mathbf{x}\|$ given alignment, $\Theta/\pi$ is the alignment probability, and $\mathbb{I}(\cdot)$ is the indicator function.

Using the probability generating functional (PGFL) of the PPP, the outage probability is
\vspace{-0.5em}
\begin{align}
\mathbb{P}_{\text{out,s}} &= \mathbb{E}_{\Phi_s} \left[ \prod_{x_i \in \Phi_s} (1 - p(x_i)) \right] \nonumber \\
&= \exp\left( -\lambda_s \int_{\mathbb{R}^2} p(\mathbf{x}) \,\mathrm{d}\mathbf{x} \right).
\end{align}
Substituting $p(\mathbf{x})$ and converting to polar coordinates yields
\begin{align}
\mathbb{P}_{\text{out,s}} &= \exp\left( -\lambda_s \int_{0}^{2\pi} \int_{0}^{R} \frac{\Theta}{\pi} \mathbb{P}_{\text{sg}}(r) \cdot r \,\mathrm{d}r \mathrm{d}\theta \right) \nonumber \\
&= \exp\left( -2\Theta\lambda_s \int_{0}^{R} r \cdot \mathbb{P}_{\text{sg}}(r) \,\mathrm{d}r \right).
\label{outs}
\end{align}
The conditional probability $\mathbb{P}_{\text{sg}}(r)$ is determined by the condition $\frac{P_{\text{ech}}}{I+N_s} > \delta$. Substituting the echo power expression:
$$
\begin{aligned}
\mathbb{P}_{\text{sg}}(r) &= \mathbb{P}\left[\frac{P \varsigma_t \varsigma_r N^2\lambda_w^2 \sigma}{(4 \pi)^3 r^{2 \alpha}} > \delta(I+N_s)\right] \\
&= \mathbb{P}\left[\sigma > \frac{\delta (4\pi)^3 r^{2\alpha}}{P \varsigma_t \varsigma_r N^2\lambda_w^2} (I+N_s)\right].
\end{aligned}
$$
Letting $\eta_s = \frac{\delta (4\pi)^3}{P \varsigma_t \varsigma_r N^2\lambda_w^2}$, and averaging over the Swerling-I randomness of $\sigma$, we obtain
$$
\begin{aligned}
\mathbb{P}_{\text{sg}}(r) &= \mathbb{E}_I\left[ \int_{\eta_s r^{2\alpha}(I+N_s)}^{\infty} \frac{1}{\bar{\sigma}} e^{-s/\bar{\sigma}} \mathrm{d}s \right] \\
&= \mathbb{E}_I\left[ \exp\left(-\frac{\eta_s r^{2\alpha}(I+N_s)}{\bar{\sigma}}\right) \right] \\
&= \exp\left(-\frac{\eta_s r^{2\alpha}N_s}{\bar{\sigma}}\right) \mathbb{E}_I\left[\exp\left(-\frac{\eta_s r^{2\alpha}}{\bar{\sigma}} I\right)\right].
\end{aligned}
$$
where $\mathcal{L}_I(s) = \mathbb{E}_I[\exp(-sI)]$ is the Laplace transform of the aggregate interference $I$. Then, the core of the proof is to compute this Laplace transform under the bounded path loss model. Letting $s(r) = \frac{\eta_s r^{2\alpha}}{\bar{\sigma}}$, the PGFL of the PPP gives:
$$
\begin{aligned}
&\mathcal{L}_I(s(r)) \\&= \mathbb{E}_{\Phi_s, \{h_i\}}\left[ \exp\left(-s(r) \sum_{x_i \in \Phi_s} P h_i \min(1, \|x_i\|^{-\alpha})\right) \right] \\
&= \exp\left(\!-\lambda_s \int_{\mathcal{A}} \left(1 - \mathbb{E}_h[\exp(-s(r) P h \min(1, r^{\!-\alpha}))]\right) \mathrm{d}\mathbf{x} \right),
\end{aligned}
$$
where $\mathcal{A}$ is the angular sector of width $2\Theta$ and $r=\|\mathbf{x}\|$. Since $h \sim \exp(1)$, we have $\mathbb{E}_h[e^{-ch}] = \frac{1}{1+c}$. Thus,
$$
\mathcal{L}_I(s(r)) = \exp\left( -\lambda_s \int_{\mathcal{A}} \frac{s(r) P \min(1, r^{-\alpha})}{1+s(r) P \min(1, r^{-\alpha})} \mathrm{d}\mathbf{x} \right).
$$
Evaluating the integral over $\mathcal{A}$, we have
\begin{align}
\int_{\mathcal{A}} (\cdot) \mathrm{d}\mathbf{x} &= \int_{-\Theta}^{\Theta} \int_{0}^{\infty} \frac{s(r) P \min(1, \varrho^{-\alpha})}{1+s(r) P \min(1, \varrho^{-\alpha})} \varrho \mathrm{d}\varrho \mathrm{d}\phi \nonumber\\
&= 2\Theta \left[ \int_{0}^{1} \!\!\frac{s(r) P}{1\!+\!s(r) P} \varrho \mathrm{d}\varrho + \int_{1}^{\infty} \!\!\!\frac{s(r) P \varrho^{-\alpha}}{1\!+\!s(r) P \varrho^{-\alpha}} \varrho \mathrm{d}\varrho \right].\nonumber
\end{align}
The first integral corresponding to near-field interferers is
\begin{align}
\frac{s(r) P}{1+s(r) P} \int_{0}^{1} \varrho \mathrm{d}\varrho = \frac{s(r) P}{2(1+s(r) P)}.\nonumber    
\end{align}
And the second integral for far-field interferers is
\begin{align}
\int_{1}^{\infty} \!\!\frac{s(r) P\varrho^{1-\alpha}}{1\!+\!s(r) P\varrho^{-\alpha}} \mathrm{d}\varrho \nonumber\!=\! \frac{s(r) P}{\alpha\!-\!2} \!\cdot \!{}_2F_1\!\left(\!1, 1\!-\!\frac{2}{\alpha}; 2\!-\!\frac{2}{\alpha};\! -s(r) P\!\right), \nonumber   
\end{align}
where ${}_2F_1(a,b;c;z)$ is the Gaussian hypergeometric function. Let $\rho(r) = s(r)P$. Combining these results provides the expression for $\mathbb{P}_{\text{sg}}(r)$. Substituting this back into Eq. \eqref{outs} and using $\mathbb{P}_{\text{cv}}^{\text{s}} = 1 - \mathbb{P}_{\text{out,s}}$ concludes the proof. \hspace*{\fill}\qed
\vspace{-0.5em}
\section{Proof of Lemma 3}
To prove \textit{Lemma 3}, we derive the DE for each component of the SINR expression in Eq. \eqref{p15}. We rely on results from large-dimensional random matrix theory, focusing on the asymptotic behavior as the number of antennas $\mathcal{\aleph} \rightarrow \infty$. For tractability, we normalize each term in Eq. \eqref{p15} by $\mathcal{\aleph}^2$.

We begin with the numerator of Eq. \eqref{p15}, representing the desired signal power. Its normalized form, denoted by $S_k$, is
\vspace{-0.5em}
\begin{align}
S_k=\frac{\mu}{\mathcal{\aleph}^2}\left|\mathbb{E}\left[\mathbf{h}_k^{\mathrm{H}} \mathbf{C}_k \hat{\mathbf{h}}_k\right]\right|^2.   
\label{p24}
\end{align}
This requires the DE of $\mu$, derived as
\vspace{-0.5em}
\begin{align}
\mu & =\frac{1}{\frac{1}{\mathcal{\aleph}} \mathbb{E}\left[\sum_{i \in \Phi_u} \hat{\mathbf{h}}_i^{\mathrm{H}} \mathbf{C}_i^2 \hat{\mathbf{h}}_i\right]} \nonumber\\
& \asymp\left(\frac{1}{\mathcal{\aleph}} \sum_{i \in \Phi_u}{\left.\operatorname{tr} \mathbf{C}_i^2 \mathbf{\Omega}_i\right)^{-1}}\right. \nonumber\\
& =\left(\frac{1}{\mathcal{\aleph}} \sum_{i \in \Phi_u}{\operatorname{tr} \mathbf{C}_i}^{-1}\right)^{-1}  =\bar{\mu},
\label{p25}
\end{align}
where the asymptotic equivalence in the second step follows from \cite{ref40}. The third step is obtained by using the identity $\mathbf{C}_i^2 \boldsymbol{\Omega}_i = \mathbf{C}_i (\mathbf{C}_i \boldsymbol{\Omega}_i) = \mathbf{C}_i (\boldsymbol{\Omega}_i^{-1} \boldsymbol{\Omega}_i) = \mathbf{C}_i$. Next, we evaluate the expectation term within $S_k$:
\begin{align}
\frac{1}{\mathcal{\aleph}} \mathbb{E}\left[\mathbf{h}_k^H \mathbf{C}_k \hat{\mathbf{h}}_k\right] & \overset{(a)}{=}\frac{1}{\mathcal{\aleph}} \mathbb{E}\left[\left(\hat{\mathbf{h}}_k^H+\tilde{\mathbf{e}}_k^H\right) \mathbf{C}_k \hat{\mathbf{h}}_k\right] \nonumber\\
& =\frac{1}{\mathcal{\aleph}} \mathbb{E}\left[\hat{\mathbf{h}}_k^H \mathbf{C}_k \hat{\mathbf{h}}_k\right] \nonumber\\
& \asymp \frac{1}{\mathcal{\aleph}} \operatorname{tr} \mathbf{C}_k \Omega_k =1.
\label{p27}
\end{align}
Step (a) holds because the channel estimate $\hat{\mathbf{h}}_k$ and the estimation error $\tilde{\mathbf{e}}_k$ are uncorrelated. And the subsequent asymptotic equivalence is an application of \cite{ref40}. Substituting these results into Eq. \eqref{p24}, the DE of the desired signal power is $
\bar{S}_k=\bar{\mu}.   
$
For the denominator of Eq. \eqref{p15}, the variance of the desired signal converges to zero asymptotically:
\begin{align}
\frac{1}{\mathcal{\aleph}^2} \operatorname{var}\left[\mathbf{h}_k^{\mathrm{H}} \mathbf{C}_k \hat{\mathbf{h}}_k\right]-\frac{1}{\mathcal{\aleph}^2} \mathbb{E}\left[\left|\tilde{\mathbf{e}}_k^{\mathrm{H}} \mathbf{C}_k \hat{\mathbf{h}}_k\right|^2\right] \xrightarrow[\mathcal{\aleph} \rightarrow \infty]{\text{a.s.}} 0,   
\label{p29}
\end{align}
which is established by using the identity $\operatorname{var}[x]=\mathbb{E}\left[|x|^2\right]-|\mathbb{E}[x]|^2$, the relation $\tilde{\mathbf{e}}_k=\mathbf{h}_k-\hat{\mathbf{h}}_k$, and applying \cite{ref40}. Applying theorem again to the remaining term yields its DE:
\begin{align}
\frac{1}{\mathcal{\aleph}^2} \mathbb{E}\left[\left|\tilde{\mathbf{e}}_k^H \mathbf{C}_k \hat{\mathbf{h}}_k\right|^2\right] & \asymp \frac{1}{\mathcal{\aleph}^2} \operatorname{tr} \mathbf{C}_k^2 \boldsymbol{\Omega}_k\left(\mathbf{\Lambda}_k-\boldsymbol{\Omega}_k\right) \nonumber\\
& =\frac{1}{\mathcal{\aleph}^2} \operatorname{tr}\left(\mathbf{D \Lambda}_k^{-1}-\mathbf{I}_{\mathcal{\aleph}}\right).
\label{p30}
\end{align}

The final term in the denominator of Eq. \eqref{p15} represents the multi-user interference. Its DE is derived as
\vspace{-0.5em}
\begin{align}
\frac{1}{\mathcal{\aleph}^2} \mathbb{E}\left[\left|\mathbf{h}_k^{\mathrm{H}} \mathbf{C}_i \hat{\mathbf{h}}_i\right|^2\right] & \asymp \frac{1}{\mathcal{\aleph}^2} \operatorname{tr} \mathbf{C}_i^2 \boldsymbol{\Omega}_i \mathbf{\Lambda}_k \nonumber\\
& =\frac{1}{\mathcal{\aleph}^2} \operatorname{tr} \mathbf{D \Lambda}_i^{-2} \mathbf{\Lambda}_k,
\label{p31}
\end{align}
which holds due to the mutual independence of $\mathbf{h}_k$ and $\hat{\mathbf{h}}_i$ for $i \neq k$. Combining Eqs. \eqref{p25}-\eqref{p31} into Eq. \eqref{p15} yields the expression in \textit{Lemma 3}. \hspace*{\fill}\qed

\vspace{-0.5em}
\section{Proof of Theorem 2}
Conditioned on the AP/user locations, the DE SINR in Eq. \eqref{p17} is expressed as
\vspace{-0.5em}
\begin{align}
\bar{\gamma}_k \asymp \frac{M N}{\frac{1}{M} \sum_{i \in \Phi_u} \sum_{m \in \Phi_c} d_m l_{m i}^{-2}\left(l_{m k}+\frac{M N}{P}\right)-1}.
\label{p32}
\end{align}
Based on this, we derive the conditional communication coverage probability $\mathbb{P}\left(\bar{\gamma}_k > T \mid \Phi_c, \Phi_u\right)$ as
\begin{align}
&\mathbb{P}\left(\bar{\gamma}_k>T \mid \Phi_c, \Phi_u\right) \nonumber \\
&=\mathbb{P}\left(\mathcal{\aleph}>T\left(\frac{1}{M} \sum_{i \in \Phi_u} \sum_{m \in \Phi_c} d_m l_{m i}^{-2}\left(l_{m k}+\frac{M N}{P}\right)-1\right)\right) \nonumber\\
& \overset{(a)}{\approx} \mathbb{P}\left(\tilde{g}>T\left(\frac{1}{M} \sum_{i \in \Phi_u} \sum_{m \in \Phi_c} d_m l_{m i}^{-2}\left(l_{m k}+\frac{M N}{P}\right)-1\right)\right) \nonumber\\
& \overset{(b)}{\approx} \!1\!-\!\!\left(\!1\!-\!\exp \!\left(\!-\!\eta_c T\!\left(\!\frac{1}{M}\! \!\sum_{i \in \Phi_u} \!\sum_{m \in \Phi_c} \!\!d_m l_{m i}^{-2}\!\left(\!l_{m k}\!+\!\frac{M\! N}{P}\!\right)\!-\!1\!\right)\!\!\right)\!\!\right)^{\!\!\tilde{\mathcal{\aleph}}}  \nonumber\\
& \overset{(c)}{=}\sum_{n=1}^{\tilde{\mathcal{\aleph}}}\binom{\tilde{\mathcal{\aleph}}}{n}(-1)^{n+1}\exp \!\bigg(\!-\!n \eta_c T \nonumber\\
&\quad \quad \quad\times\left.\!\!\left(\frac{1}{M}\! \sum_{i \in \Phi_u}\! \sum_{m \in \Phi_c} \!\!d_m l_{m i}^{-2}\!\left(\!l_{m k}\!+\!\frac{M N}{P}\right)\!-1\!\right)\!\right)\!\label{p36}, 
\end{align}
where in step (a), we approximate the aggregate array gain $\mathcal{\aleph}$ with a Gamma-distributed variable $\tilde{g}$ having shape parameter $\tilde{\mathcal{\aleph}}=\mathbb{E}[\mathcal{\aleph}]$, a tight approximation for massive MIMO. In step (b), we invoke Alzer's inequality \cite{ref41} to bound the Gamma CCDF, where $\eta_c=\tilde{\mathcal{\aleph}}(\tilde{\mathcal{\aleph}}!)^{-\frac{1}{\tilde{\mathcal{\aleph}}}}$. In step (c), it is expanded using the binomial theorem for analytical convenience.

To obtain $\mathbb{P}_{\text{cv}}^{\text{c}}$, we take the expectation of Eq. \eqref{p36} over the spatial processes. Let $\mathcal{I}_{m k} = d_m l_{m i}^{-2}\left(l_{m k}+\frac{M N}{P}\right)$. Applying Jensen's inequality yields a tractable lower bound:
\vspace{-0.5em}
\begin{align}
\mathbb{P}_{\text{cv}}^{\text{c}}= & \sum_{n=1}^{\tilde{ \aleph }}\binom{\tilde{ \aleph }}{n}(-1)^{n+1} \nonumber\\
& \times \mathbb{E} \left[\exp \left(-n \eta_c T\left(\frac{1}{M} \sum_{i \in \Phi_u} \sum_{m \in \Phi_c} \mathcal{I}_{m k}-1\right)\right)\right] \nonumber\\
\geq & \sum_{n=1}^{\tilde{ \aleph }}\binom{\tilde{ \aleph }}{n}(-1)^{n+1}  \nonumber\\
& \times \exp \left(-n \eta_c T \mathbb{E} \left[\frac{1}{M} \sum_{i \in \Phi_u} \sum_{m \in \Phi_c} \mathcal{I}_{m k}-1\right]\right).
\label{p38}
\end{align}
We evaluate the expectation using the law of large numbers. As the network area radius $R \rightarrow \infty$, the ratio of users to APs converges to $\lambda_u/\lambda_c$:
\vspace{-0.5em}
\begin{align}
\lim_{R \rightarrow \infty} \mathbb{E}\left[\frac{1}{M} \sum_{i \in \Phi_u} \sum_{m=1}^M \mathcal{I}_{m k}\right]
=\frac{\lambda_u}{\lambda_c} \mathbb{E}\left[\mathcal{I}_{m k}\right],
\label{p41}
\end{align}
Substituting $d_m $ into $\mathcal{I}_{mk}$ reveals two components:
\vspace{-0.5em}
\begin{align}
\mathcal{I}_{mk} =& \left( \sum_{j \in \Phi_u} l_{mj} |\boldsymbol{\psi}_j^{\mathrm{H}} \boldsymbol{\psi}_k|^2 + \frac{1}{\tau_{\mathrm{tr}}\rho_{\mathrm{tr}}} \right) l_{mi}^{-2} l_{mk} \nonumber\\&+ \left( \sum_{j \in \Phi_u} l_{mj} |\boldsymbol{\psi}_j^{\mathrm{H}} \boldsymbol{\psi}_k|^2 + \frac{1}{\tau_{\mathrm{tr}}\rho_{\mathrm{tr}}} \right) l_{mi}^{-2} \frac{MN}{P}.
\end{align}
The pilot contamination term $|\boldsymbol{\psi}_j^{\mathrm{H}} \boldsymbol{\psi}_k|^2$ acts as a Bernoulli variable that equals $1$ (collision) with probability $1/\tau_{\mathrm{tr}}$ and $0$ otherwise, effectively thinning $\Phi_u$ into a process with intensity $\lambda_u' = \lambda_u / \tau_{\mathrm{tr}}$. Applying Campbell's theorem to this thinned process, we decompose the total expectation into $\mathcal{I}_1$ and $\mathcal{I}_2$:
\vspace{-0.5em}
\begin{align}
\mathcal{I}_1 = \frac{\lambda_u}{\lambda_c} \left( \frac{\lambda_u}{\tau_{tr}} \mathbb{E} \left[ l_{m j} l_{m i}^{-2} l_{m k}\right] + \frac{1}{\tau_{tr} \rho_{tr}} \mathbb{E} \left[l_{m i}^{-2} l_{m k}\right] \right),
\label{p42}
\end{align}
and
\vspace{-0.5em}
\begin{align}
\mathcal{I}_2= \frac{\tilde{\mathcal{\aleph}}}{P} \frac{\lambda_u}{\lambda_c} \left( \frac{\lambda_u}{\tau_{tr}} \mathbb{E}\left[ l_{m j} l_{m i}^{-2} \right] + \frac{1}{\tau_{\mathrm{tr}} \rho_{\mathrm{tr}}} \mathbb{E}\left[ l_{m i}^{-2} \right] \right).
\label{p43}
\end{align}
The derivation of these terms involves computing expectations over various products and ratios of path-loss terms. We utilize the independence of channel components and apply Jensen's inequality to obtain lower bounds. For the cross-product terms in $\mathcal{I}_1$, we have
\vspace{-0.5em}
\begin{align}
\mathbb{E}\left[l_{m j} l_{m i}^{-2} l_{m k}\right] & =\mathbb{E}\left[l_{m j}\right] \mathbb{E}\left[l_{m i}^{-2}\right] \mathbb{E}\left[l_{m k}\right] \nonumber\\
& \geq \mathbb{E}\left[l_{m j}\right] \mathbb{E}\left[l_{m i}^{-1}\right]^2 \mathbb{E}\left[l_{m k}\right] \geq 1.\nonumber
\end{align}

We now evaluate the inverse path loss terms. The expectation $\mathbb{E}[l_{mi}^{-v}]$ is bounded using Jensen's inequality:
\vspace{-0.5em}
\begin{align}
\mathbb{E}\left[l_{m i}^{-v}\right] & =\mathbb{E}\left[\frac{1}{l_{m i}^v}\right] \geq \frac{1}{\mathbb{E}\left[l_{m i}^v\right]},
\label{p53}
\end{align}
where $\mathbb{E}\left[l_{m i}^v\right]$ is calculated using the path-loss model as
\vspace{-0.5em}
\begin{align}
\mathbb{E}\left[l_{m i}^v\right]  =2 \pi\left(\int_0^1 y \mathrm{~d} y+\!\int_1^{\infty} \!\!y^{-v \alpha+1} \mathrm{~d} y\right) =\frac{v \alpha \pi}{v \alpha-2}.
\label{p55}
\end{align}
Consequently, we have
\vspace{-0.5em}
\begin{align} \mathbb{E}[l_{mj} l_{mi}^{-2}] &= \mathbb{E}[l_{mj}]  \mathbb{E}[l_{mi}^{-2}] \nonumber \\ & \geq \mathbb{E}[l_{mj}]  \frac{1}{\mathbb{E}[l_{mi}^2]} \nonumber \\  &= \frac{\alpha-1}{\alpha-2}.
\label{p60}
\end{align}
Substituting these moments back into Eqs. \eqref{p42}-\eqref{p43} yields:
\begin{align}
\mathcal{I}1 &= \frac{\lambda_u}{\lambda_c \tau_{tr}} \left( \lambda_u +  \frac{\alpha-2}{\rho_{tr}\alpha \pi} \right), \label{p61} \\
\mathcal{I}2 &= \frac{\lambda_u \tilde{\mathcal{\aleph}}}{\lambda_c \tau_{tr} P} \left( \frac{\alpha-1}{\alpha-2}\lambda_u + \frac{\alpha-1}{\rho_{tr}\alpha \pi} \right). \label{p62}
\end{align}
Substituting $\mathcal{I}_1$ and $\mathcal{I}_2$ into Eq. \eqref{p41} and then into Eq. \eqref{p38} completes the proof. \hspace*{\fill}\qed

\end{document}